\documentclass[11pt]{article}
\pdfoutput=1
\usepackage{jheppub}
\usepackage{amsfonts}
\usepackage{amsmath}
\usepackage{mathtools}
\allowdisplaybreaks[4]         
\usepackage{amssymb}   
\usepackage{euscript}       
\usepackage[dvipsnames]{xcolor}
\usepackage{IEEEtrantools}
\usepackage{upgreek}
\usepackage{graphicx}
\usepackage{caption}
\usepackage{subcaption}
\usepackage[export]{adjustbox}
\usepackage[makeroom]{cancel}
\usepackage{physics}
\usepackage{tensor}
\usepackage{braket}
\usepackage{float}
\usepackage{color}
\usepackage[normalem]{ulem}
\usepackage{starfont}

\usepackage{hyperref}
\hypersetup{
	colorlinks =WildStrawberry,
	linkcolor =red,
	pdftitle={Disks globally maximize the entanglement entropy in $2+1$ dimensions},
	citecolor=MidnightBlue,
	filecolor=WildStrawberry,
	urlcolor=WildStrawberry,
}

\newcommand{\req}[1]{(\ref{#1})} %{Eq.\thinspace(\ref{#1})}

\newcommand{\rd}[1]{{\color{black}#1}}
\newcommand{\be}{\begin{equation}}
\newcommand{\ee}{\end{equation}}

\newcommand{\diff}{\mathrm{d}}

\newcommand{\ssc}{\scriptscriptstyle}
\newcommand{\eg}{{\it e.g.,}\ }
\newcommand{\ie}{{\it i.e.,}\ }

\newcommand{\SEMI}{S_{\text{EE}}^{\rm \ssc EMI}}

\newcommand{\see}{S_{\text{EE}}}

\makeatletter
\newcommand{\dal}{\mathop{\mathpalette\dal@\relax}}
\newcommand{\dal@}[2]{%
  \begingroup
  \sbox\z@{$\m@th#1\square$}%
  \dimen0=\fontdimen8
    \ifx#1\displaystyle\textfont\else
    \ifx#1\textstyle\textfont\else
    \ifx#1\scriptstyle\scriptfont\else
    \scriptscriptfont\fi\fi\fi3
  \makebox[\wd\z@]{%
    \hbox to \ht\z@{%
      \vrule width \dimen0
      \kern-\dimen0
      \vbox to \ht\z@{
        \hrule height \dimen0 width \ht\z@
        \vss
        \hrule height 2\dimen0
      }%
      \kern-2.5\dimen0
      \vrule width 2.5\dimen0
    }%
  }%
  \endgroup
}
\makeatother

\title{\huge Disks globally maximize the entanglement entropy in $2+1$ dimensions}

\author[\text{\Zeus}]{Pablo Bueno,} 
\author[\text{\Zeus}]{Horacio Casini,} 
\author[\text{\Poseidon}]{Oscar Lasso Andino} 
\author[\text{\Hades},\text{\Kronos}]{and Javier Moreno}

\affiliation[\text{\Zeus}]{Instituto Balseiro, Centro At\'omico Bariloche,\\
8400-S.C. de Bariloche, R\'io Negro, Argentina}
\affiliation[\text{\Poseidon}]{Escuela de Ciencias F\'isicas y Matem\'aticas, Universidad de Las Am\'ericas,\\
C/. Jos\'e Queri, C.P. 170504, Quito, Ecuador}
\affiliation[\text{\Hades}]{Instituto de F\'isica, Pontificia Universidad Cat\'olica de Valpara\'iso,\\ Casilla 4059, Valpara\'iso, Chile.}
\affiliation[\text{\Kronos}]{Center for Quantum Mathematics and Physics (QMAP),\\
Department of Physics \& Astronomy, University of California, Davis, CA 95616 USA}

\vspace{0.3cm}
\emailAdd{pablo.bueno@cab.cnea.gov.ar, casini@cab.cnea.gov.ar, oscar.lasso@udla.edu.ec, francisco.moreno.g@mail.pucv.cl} 

\abstract{The entanglement entropy corresponding to a smooth region in general three-dimensional CFTs contains a constant universal term, $  -F \subset \see$. For a disk region, $F|_{\rm disk}\equiv F_0$ coincides with the free energy on $\mathbb{S}^3$ and provides an RG-monotone for general theories. As opposed to the analogous quantity in four dimensions, the value of $F$ generally depends in a complicated (and non-local) way on the geometry of the region and the theory under consideration. 
For small geometric deformations of the disk in general CFTs as well as for arbitrary regions in holographic theories, it has been argued that $F$ is precisely minimized by disks. Here, we argue that $F$ is globally minimized by disks with respect to arbitrary regions and for general theories. The proof makes use of the strong subadditivity of entanglement entropy and the geometric fact that one can always place an osculating circle within a given smooth entangling region. For topologically non-trivial entangling regions with $n_B$ boundaries, the general bound can be improved to $F \geq n_B F_0$. In addition, we provide accurate approximations to $F$ valid for general CFTs in the case of elliptic regions for arbitrary values of the eccentricity which we check against lattice calculations for free fields.  We also evaluate $F$ numerically for more general shapes in the so-called ``Extensive Mutual Information model'', verifying the general bound.       }

\begin{document}

\maketitle  

%\newpage
\section{Introduction}
\begin{figure}[t] \centering
	\includegraphics[scale=0.35]{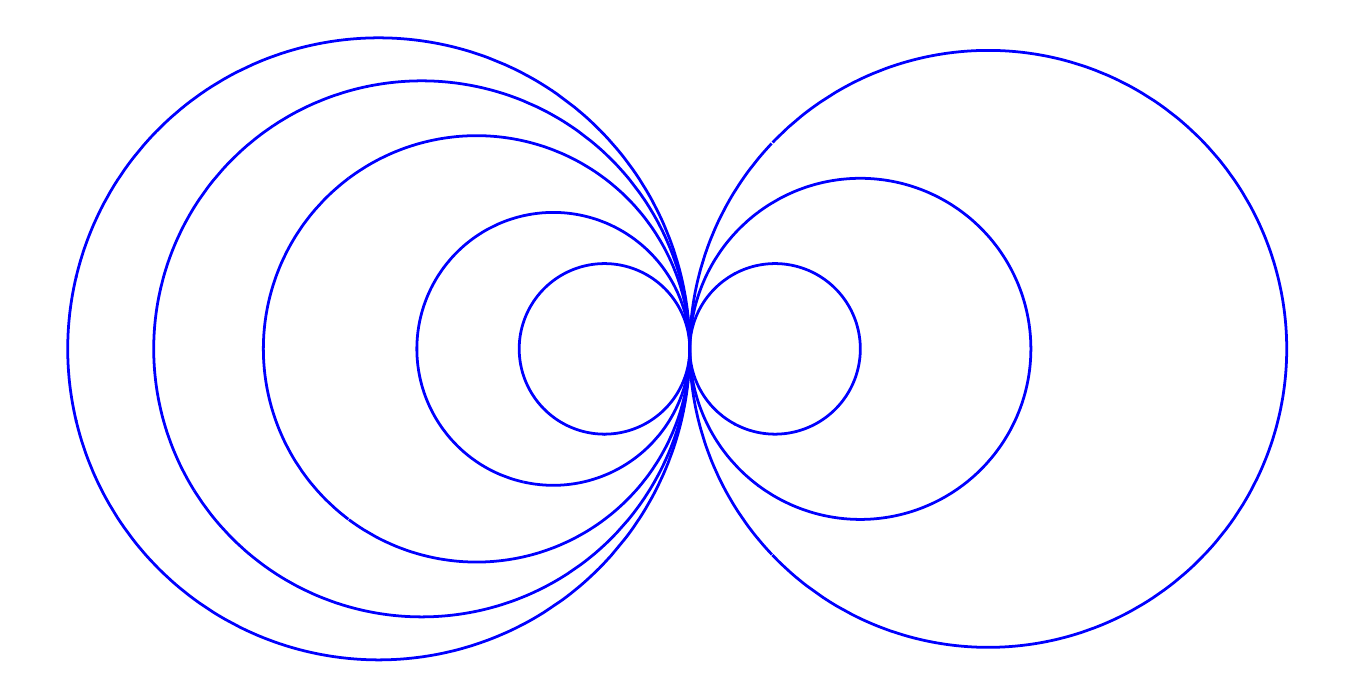}\hspace{0.2cm}	
	\includegraphics[scale=0.4]{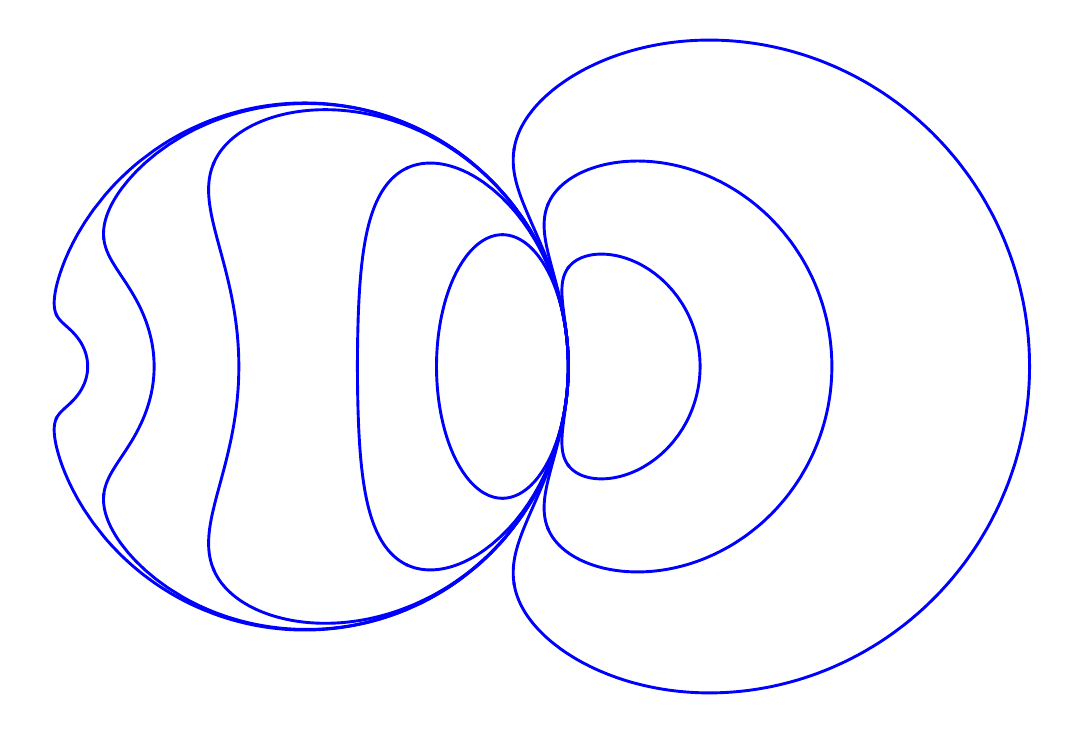}\hspace{0.2cm}
	\includegraphics[scale=0.37]{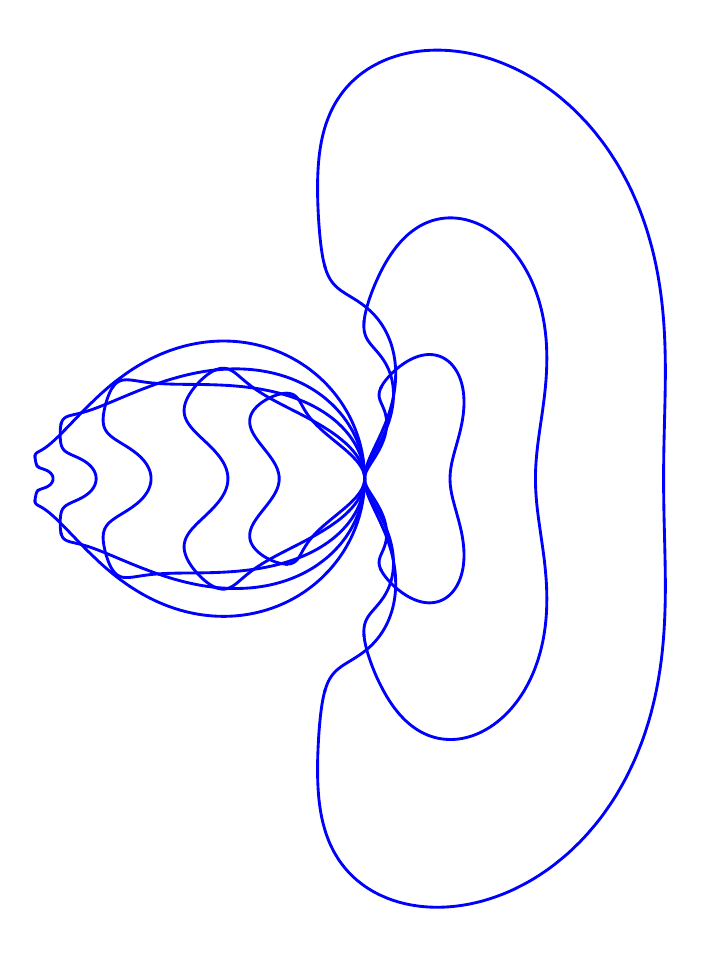}
	\caption{ \textsf{We show three sets of entangling regions (left, middle and right plots, respectively). In each set, all the different regions are related to each other through conformal transformations (the same transformations are applied in each set), and therefore share the same $F$ for a given CFT. Disks are naturally unchanged, but notice for instance how an ellipse (innermost figure in middle plot) can be deformed, without modifying $F$, into a disk with a small depression.} }
	\label{refiss376}
\end{figure}

In spite of its intrinsically divergent nature, the entanglement entropy (EE) of spatial regions in $d$-dimensional CFTs contains pieces which are well defined in the sense of being independent of the regularization utilized. The nature of such ``universal'' terms changes considerably depending on whether $d$ is even or odd. In even dimensions, the prototypical universal contributions for smooth surfaces are coefficients of  logarithmic terms and have an intrinsically UV origin. In particular, they are local in the entangling surface $\partial A$, meaning that they are given by integrals over such surface of various intrinsic and extrinsic curvatures. These appear weighted by certain theory-dependent constants which are proportional to the corresponding trace-anomaly coefficients  \cite{Solodukhin:2008dh,Fursaev:2012mp,Safdi:2012sn,Miao2015a}. In odd dimensions, there is no logarithmic term for smooth regions\footnote{In this paper we will not be concerned with singular entangling regions. In that case, a logarithmic divergence arises ---see \eg \cite{Bueno:2019mex}--- which always reduces the EE more than any smooth region. Hence, in the task of characterizing regions which maximize EE, such regions can be discarded from the beginning.  } and the universal piece corresponds to the constant term, independent of the cutoff. As opposed to logarithmic terms, constant terms are not given by local integrals over the entangling surface. Rather, they generally depend in a complicated way on the theory under consideration as well as  on the shape of the entangling region $A$. The simplest case corresponds to three-dimensional CFTs. For those, the EE of a smooth region takes the form
\begin{equation}\label{entro}
\see(A) = c_0 \frac{{\rm perimeter} (\partial A)}{\delta} - F(A)+\mathcal{O}(\delta)\, ,
\end{equation}
where $c_0$ is a non-universal constant and $\delta$ is a UV regulator. On the other hand, $F(A)$ is a universal, dimensionless, and generally non-local in nature piece. 
%The first term, proportional to the boundary perimeter, is divergent and non-universal, while $F(A)$ is universal, dimensionless, and generally non local in nature. 
Importantly, $F(A)$ is a conformal invariant quantity, so that $F(A)=F\left(T(A) \right)$, where $T$ is any conformal transformation ---see Fig.\,\ref{refiss376}.

When $A$ is a disk ---namely, a region with $\partial A = \mathbb{S}^1$--- the corresponding $F$, which henceforth we denote $F_0$, equals the free energy of the corresponding theory on $\mathbb{S}^3$ \cite{Dowker:2010yj,CHM}. $F_0$ turns out to define an RG monotone for general quantum field theories, hence providing a version of the $F$-theorem in three dimensions \cite{Casini:2012ei}.
%\comment{important role in the proof of the F-theorem}\\

In this paper we would like to ask the question of which region $A$ minimizes the universal part of the entanglement entropy in three dimensions for a given CFT. Naturally, the obvious candidate corresponds to the disk region, \ie 
\begin{equation}\label{Fiso}
F(A)/F_0 \geq 1\, , \quad  \text{with} \quad F(A)/F_0=1 \Leftrightarrow  A = \text{disk}\, .
\end{equation}
For small geometric deviations of the disk, the above inequality follows from Mezei's formula \cite{Mezei:2014zla} ---see \req{fmeze0} below--- which informs us that corrections to $F_0$ are given by a positive-definite geometry-dependent coefficient weighted by the (also positive-definite) stress-tensor coefficient\footnote{It is worth pointing out that in the case of the Euclidean free energy, $\mathbb{S}^3$ also provides a local maximum and that the leading correction to the free energy for slightly deformed spheres is similarly controlled by $C_{\ssc T}$ \cite{Bobev:2017asb,Fischetti:2017sut}. Beyond the perturbative regime, partial results for free fermions suggest that $\mathbb{S}^3$ does not necessarily correspond to a global maximum  \cite{Bobev:2017asb}. However, the match between the EE and the Euclidean free energy does not go through beyond the  round case, so we cannot extrapolate such results to the present context. For additional related results regarding the extremization of three-dimensional free energies in Lorentzian backgrounds, see \cite{Fischetti:2018shp,Fischetti:2020knf,Cheamsawat:2020awh}. } $C_{\ssc T}$. On general grounds, one can also argue that $F$ tends to grow as regions incorporate sufficiently thin sectors, as in that case $F$ includes contributions proportional to the ratio between the corresponding long and short dimensions ---see \eg \cite{Casini:2005zv,Ryu:2006ef}. As far as particular theories are concerned, the only model for which \req{Fiso} has a definite positive resolution is holographic Einstein gravity. In that case, $F^{\rm holo}$ can be written in terms of the Willmore energy  \cite{marques2014willmore,willmore1996riemannian,toda} of a duplicated version of the corresponding Ryu-Takayanagi surface \cite{alexakis2008renormalized,Astaneh:2014uba,Fonda:2015nma} and the global minimization of $F^{\rm holo}$ by $F^{\rm holo}_0$ follows from simple geometric arguments. 

An analogous problem has been studied for the universal (logarithmic) contribution to the EE in four-dimensions in \cite{Astaneh:2014uba} \rd{ and \cite{Perlmutter:2015vma}}. In that case, however, the essentially local nature of the corresponding term makes it be fully determined in terms of two (theory-independent) local integrals over $\partial A$ weighted by the trace-anomaly charges of each theory \rd{$a$ and $c$. The global minimization  of the EE universal term  by the round sphere follows then rather straightforwardly for general CFTs in the case of arbitrary entangling surfaces of genus $g=0$ and $g=1$. On the other hand, as pointed out in \cite{Perlmutter:2015vma}, when $a>c$ it can be shown that the EE universal coefficient is actually unbounded from below for sufficiently high genus.\footnote{ \rd{The argument follows from two facts: i) the EE universal coefficient can be written as a linear combination of the Willmore functional of $\partial A$ plus $(a/c-1)$ times the Euler characteristic of $\partial A$ ---which is proportional to $(1-g)$, where $g$ is the genus of the entangling surface; ii)  for every genus $g$ there exists at least a surface whose Willmore energy lies somewhere between the values $4\pi$ and $8\pi$  \cite{10.2307/1970625,Kusner}. Then, whenever $(a/c-1)>0$, a growing $g$ will make the Euler characteristic piece more and more negative, whereas the Willmore part for certain geometries will remain bounded between the aforementioned values independently of $g$.} } }

% and the global extremization of EE by the round sphere follows straightforwardly for general CFTs.%\footnote{It is important for this to work the fact that the piece proportional to the $c$-charge is a positive semi-definite combination of extrinsic curvatures. This implies that global minimization of the full combination is realized for the surface with the smallest Euler characteristic for which such piece vanishes, which is the sphere.}

Here, we will argue that disks globally maximize the EE for general CFTs in three dimensions. Before getting to the proof, we start with a brief summary of previous results and general remarks in Section \ref{previous}. Then, in Section \ref{elipsq} we study in detail the case of elliptic regions. In particular, we provide analytic approximations of $F$ for such regions for arbitrary values of the eccentricity for general CFTs. These we compare with the exact result obtained numerically for the so-called ``Extensive Mutual Information'' (EMI) model as well as with lattice results for free scalars and fermions finding good agreement for the latter and a small puzzle for the former. The lattice results are obtained using mutual information (MI) as a geometric regulator which requires the evaluation of the MI for two concentric regions and the subtraction of a purely geometric piece weighted by the coefficient controlling the EE of a thin strip region. In Section \ref{secemi} we evaluate  $F$ for several families of more general entangling regions in the EMI model. This allows to illustrate to what extent $F$ may vary in general as the geometric features of the entangling region non-perturbatively deviate from the most symmetric cases. Naturally, all results in these two sections are in agreement with \req{Fiso}, as expected. In Section \ref{proof} we prove \req{Fiso} for general theories. In order to do that, we first study the effect on the EE of considering geometric perturbations of the entangling region for which the derivatives of the perturbation go to zero with a different velocity than the perturbation itself and characterize the type of deformations which allow us to to separate the leading correction to $F$ from the subleading ones. Denoting by  $\gamma(s)$ the curve describing the entangling surface $\partial A$ and $\eta(s)$ its normal vector at a point $s$, the relevant deformations $\gamma_{\lambda}(s)=\gamma(s)+\delta_{\lambda}(s) \eta(s)$ are such that their $k$-th derivative behaves as $|| \delta_{\lambda}^{(k)}|| \sim \lambda^{4-k}$ with $\lambda\rightarrow 0$, where $||\cdot||$ is the uniform norm. We then show that any functional allowing for a hierarchy of perturbative terms in $\delta_{\lambda}$ in the sense just described which is  Euclidean invariant, strong superadditive and constant for disks (which includes $F$ as a particular case) is globally minimized for disks. Three facts play a crucial role in the proof: i) for a given entangling region which fully contains another one and such that both coincide within a given interval, functionals of that class always increase more for the outer region than for the inner one as we perturb both outwards at the point of osculation; ii) disks have a vanishing first-order correction to $F$; iii) it is always possible to place an osculating disk within a given region.     We conclude in Section \ref{finalc} with some final comments. In particular, we show there that our results in Section  \ref{secemi} imply a stronger bound for topologically non-trivial regions ---\ie regions with holes and/or multiply connected subregions. In particular, we find $F \geq n_B  F_0$, where $n_B$ is the number of boundaries of the region. In Appendix \ref{EMIMIregu} we explore how the MI regularization of the EE works in the case of the EMI model, for which we have full control of the calculations. In particular, we study how the practical limitations of the lattice involving the finiteness of the regions affect the results and extract some conclusions which we use in Section \ref{elipsq}.

\section{General observations and previous results}\label{previous}
In this section we review what is known about the issue of whether or not $F$ is minimized by disk regions. The evidence for general CFTs is essentially limited to small perturbations of the disk and to regions which contain very thin sectors. On the other hand, in the particular case of holographic theories dual to Einstein gravity, $F$ can be written in terms of the so-called Willmore energy of a duplicated version of the corresponding Ryu-Takayanagi surface. This in turn implies \req{Fiso} for holographic Einstein gravity. 

\subsection{Perturbative deformations of the disk for general CFTs}
The question of whether or not $F$ is minimized for a disk region for arbitrary CFTs has a positive answer when the analysis is restricted to figures which are small deformations of it.
In particular, for a slightly deformed disk defined by the polar equation
\begin{equation}\label{fiy}
r(\theta)=R \left[1+ \epsilon \sum_{\ell} \left( \frac{a^{(c)}_{\ell}}{\sqrt{\pi}} \cos (\ell \theta) + \frac{a^{(s)}_{\ell}}{\sqrt{\pi}} \sin (\ell \theta) \right) \right]\, , \quad \epsilon \ll 1\, ,
\end{equation} 
(where $a^{(c)}_{\ell}$, $a^{(s)}_{\ell}$ are some numerical coefficients which determine the shape of the perturbation),
the result for $F$ at leading order in $\epsilon$ is given, for general CFTs, by the three-dimensional version of Mezei's formula \cite{Mezei:2014zla,Faulkner:2015csl}
\begin{equation}\label{fmeze0}
F=F_0+ \epsilon^2 \, \frac{\pi^3 C_{\ssc T} }{24} \sum_{\ell} \ell (\ell^2-1)
\left[ (a^{(c)}_{\ell})^2+(a^{(s)}_{\ell})^2 \right] + \mathcal{O}(\epsilon^4) \, , \end{equation}
where $C_{\ssc T}$ is the coefficient which controls the flat-space stress-tensor two-point function charge.\footnote{This is defined from: \cite{Osborn:1993cr} $\braket{T_{ij}(x)T_{kl}(0)}_{\mathbb{R}^d}=C_{\ssc T}\tfrac{1}{x^{2d}}\left[I_{i(k}I_{l)j}-\frac{\delta_{ij}\delta_{kl}}{d}\right]$, where $I_{ij}\equiv \delta_{ij}-2\frac{x_i x_j}{x^2}$ is a theory-independent structure.} In this expression, everything is determined by the geometry of the entangling region with the exception of $C_{\ssc T}$, which is the only theory-dependent piece. This is a positive-definite quantity for unitary CFTs and hence it is evident from \req{fmeze0} that \req{Fiso} holds, since the geometric coefficient is also positive semi-definite in all cases.

\subsection{Very thin regions for general CFTs}\label{stripi}
When one of the dimensions of the entangling region becomes very thin compared to the other, the universal contribution approaches the one corresponding to a strip ---whose boundary is defined by two parallel straight lines of length $L$ and separated a distance $r\ll L$. In that situation, $F$ tends to diverge as
\begin{equation}\label{strip}
F \simeq k^{(3)} \frac{L}{r} +\dots
\end{equation}
where $k^{(3)}$ is a positive coefficient characteristic of the CFT under consideration. This quantity is known, for instance, for: holographic theories \cite{Ryu:2006ef,Liu:1998bu}, free scalars and fermions \cite{Casini:2009sr}, as well as for the EMI model we consider below \cite{Bueno1}. As opposed to other charges appearing in the entanglement entropy of symmetric regions in various dimensions, $k^{(3)}$ does not have an alternative CFT interpretation beyond entanglement entropy.  The behavior \req{strip} shows that the conjectural properties \req{Fiso} are clearly satisfied for regions which possess a dimension sufficiently thinner than the other. Besides, regions which may posses various features including some which approach the strip-like shape just described will tend to make $F$ grow.

\subsection{Holography and Willmore energy}
For holographic theories dual to Einstein gravity, the entanglement entropy for a given boundary region can be obtained from the Ryu-Takayanagi prescription \cite{Ryu:2006bv,Ryu:2006ef,Hubeny:2007xt,Lewkowycz:2013nqa,Faulkner:2013ana}. The purely geometric nature of the formula has allowed for the evaluation of entanglement entropy for regions beyond the most symmetric instances. This has been done in the three-dimensional case \eg in \cite{Fonda:2014cca,Fonda:2015nma}, using the ``Surface Evolver'' tool \cite{doi:10.1080/10586458.1992.10504253} ---see also \cite{Dorn:2016eai,Katsinis:2019lrh}. As far as $F^{\rm holo}$ is concerned, it can be proven that it is indeed globally minimized by disk regions.
%The argument of the previous subsection proves \req{Fiso} for general CFTs as long as we restrict ourselves to perturbative deformations of the disk. On the other hand, if one considers the particular case of holographic theories dual to Einstein gravity, it is possible to show that 
%$F$ is indeed minimized by the disk region with respect to arbitrary (non-perturbative) shapes. 
The argument is as follows. 
%from the fact that $F^{\rm \ssc holo}$ can be written in terms of the Willmore energy of a\cite{alexakis2008renormalized,Fonda:2015nma}. 

Given a general smooth, orientable, closed surface, $\Sigma$  in $\mathbb{R}^3$ (not necessarily minimal), the so-called ``Willmore energy'' is defined as  \cite{marques2014willmore,willmore1996riemannian,toda}
\begin{equation}\label{WillmoreDef}
\mathcal{W}\left(\Sigma\right)=\int_\Sigma H^2\diff S\, ,
\end{equation}
where $H\equiv (k_1+k_2)/2$ is the mean curvature constructed from the principal curvatures  $k_1$ and $k_2$ of $\Sigma$ and $\diff S$ the surface element. This quantity has a global lower bound
\begin{equation}\label{WillmoreBound}
\mathcal{W}\left(\Sigma\right)\geq4\pi \, ,
\end{equation}
which is saturated when $\Sigma$ is a spherical surface.\footnote{If the surfaces taken into account have greater genus, the bound is higher \cite{fern2012minmax}.} In order to see why this is the case, one can modify the Willmore functional \eqref{WillmoreDef} subtracting the Gaussian curvature $K=k_1k_2$ as \cite{willmore1996riemannian}
\begin{equation}\label{WillmoreDef2}
\tilde{\mathcal{W}}\left(\Sigma\right)=\int_\Sigma \left(H^2-K\right)\diff S=\mathcal{W}\left(\Sigma\right)-\int_\Sigma K\diff S .
\end{equation}
The study of both versions of the functional is equivalent. This is because,  from the Gauss-Bonnet theorem $\int_\Sigma K\diff S=2\pi \chi\left(\Sigma\right)$, where $\chi\left(\Sigma\right)$ is the Euler characteristic of the surface. Hence, for fixed genus, $\tilde{\mathcal{W}}$ is just a constant shift  with respect to $\mathcal{W}$. The interest of the modified functional $\tilde{\mathcal{W}}$ is manifest when written in terms of the principal curvatures
\begin{equation}
\tilde{\mathcal{W}}\left(\Sigma\right)=\frac{1}{4}\int_\Sigma \left(k_1-k_2\right)^2\diff S \, . 
\end{equation}
It immediately follows that $\tilde{\mathcal{W}}\left(\Sigma\right)$ has a sign, $\tilde{\mathcal{W}}\left(\Sigma\right)\geq 0$, and also that the bound is exclusively saturated for a totally umbilical surface, $k_1=k_2$, which in the present context is just a fancy name for a sphere \cite{willmore1996riemannian}. 
%as the minimal value of $\tilde{\mathcal{W}}\left(\Sigma\right)=0$ is found when the surface is totally umbilic, $k_1=k_2$. Thus, the sphere, $\mathbb{S}^2$, is the only surface that satisfies the bound \cite{willmore1996riemannian}.
After this discussion, it is straightforward to check that the bound for the original functional \eqref{WillmoreBound} is found from \eqref{WillmoreDef2} particularizing the Euler characteristic to the  sphere case, $\chi\left(\mathbb{S}^2\right)=2$.

Now, in \cite{alexakis2008renormalized,Astaneh:2014uba,Fonda:2015nma}, it was pointed out that the finite term in the three-dimensional holographic entanglement entropy can be written in terms of a Willmore energy as
\begin{equation}\label{WillmoreFholo}
F^{\text{holo}}(A)=\frac{L_{\star}^2}{8G}\mathcal{W}\left(2\Sigma_{\mathrm{RT}(A)}\right)\, ,
\end{equation}
where $L_{\star}$ is the AdS radius and $G$ is the Newton constant ---see also \cite{Anastasiou:2020smm,Anastasiou:2018mfk,Anastasiou:2021swo,Taylor:2020uwf}. In this expression, $2\Sigma_{\mathrm{RT}(A)}\equiv \Sigma_{\mathrm{RT}(A)}\cup\Sigma_{\mathrm{RT}(A)}'$ is a closed surface embedded in $\mathbb{R}^3$ which consists of a duplicated Ryu-Takayanagi surface \cite{Babich:1992mc}: the surface $\Sigma_{\mathrm{RT}(A)}'$ is just $\Sigma_{\mathrm{RT}(A)}$ reflected with respect to the AdS boundary and joined along the entangling surface $\partial A$. 
%Relation \eqref{WillmoreFholo} was also derived in the context of renormalized entanglement entropy following the extrinsic counterterms scheme \cite{Anastasiou:2020smm}. This prescription, alternative to standard holographic renormalization \cite{Emparan:1999pm,Kraus:1999di,Balasubramanian:1999re,deHaro:2000vlm}, has been particularly useful to renormalize entanglement entropy \cite{Anastasiou:2017xjr,Anastasiou:2018mfk,Anastasiou:2021swo} and extract information of the theory, such as, the local and non-local degrees of freedom present in the dual CFT. 
Then, as mentioned earlier, the Willmore energy of the spherical surface is minimal, saturating the bound \eqref{WillmoreBound}. Hence, using the fact that the Ryu-Takayanagi surface for a disk region is an hemi-sphere, this is translated into a global bound for $F^{\mathrm{holo}}$,
\begin{equation}
F^{\mathrm{holo}}(A)\geq\frac{\pi L_{\star}^2}{2G}\, , \quad  \text{with} \quad F^{\mathrm{holo}}(A)=\frac{\pi L_{\star}^2}{2G} \Leftrightarrow A = \text{disk}\, .
\end{equation}
%From \eqref{F2} we see that the minimal bound for $F$ is translated into a maximal one for entanglement entropy, i.e., the circular entangling surface  is the most entangled one among all the possible shapes within the same topological class.

\section{Elliptic entangling regions}\label{elipsq}
Perhaps the simplest (non-perturbative) generalizations of the disk region which come to mind are ellipses. In spite of this, there do not seem to be too many results available in the literature. In the holographic context, they have been studied numerically alongside more general regions in \cite{Fonda:2014cca}. In the same context, some bounds for their corresponding $F$ were obtained in  \cite{Allais:2014ata}. In this section we start by considering the limit in which the elliptic regions are very squashed (eccentricity $e\rightarrow 1$), which leads to a general approximation in terms of the strip coefficient $k^{(3)}$ valid for general theories. Then, using this approximation along with Mezei's formula  for the opposite limit of almost round ellipses, we build a trial function which approximates $F_{(e)}$ for arbitrary values of the eccentricity for a general CFT as long as we know the values of the three coefficients $F_0$, $C_{\ssc T}$, $k^{(3)}$. We present the resulting approximations for the EMI model, holographic Einstein gravity, free scalars and free fermions. For the last two, we perform lattice calculations of  $F_{(e)}$ using MI as a regulator for $e=\sqrt{3}/2\simeq 0.866 $, $e=2 \sqrt{2}/3\simeq 0.943$, $e= \sqrt{15}/4\simeq 0.968$ and $e=2\sqrt{6}/5 \simeq 0.98$ which we compare to the approximations. The fermion results agree reasonably well with the trial function, which also approximates the exact EMI model results with a discrepancy lower than a $\sim 5 \%$ for the whole range. On the other hand, the lattice produces results for the scalar which are considerably lower than the ones obtained from the approximation. While this appears to be an artifact of the lattice results (at least to some extent) we introduce an additional trial function which does a better job in approximating the scalar results. Along with the initial one, these two functions confidently bound the possible values of $F_{(e)}$ for a given CFT.  In all cases, $F_{(e)}$ is a monotonically increasing function of the eccentricity.

 %The first of these functions approximates well the exact EMI model results for the whole range, the greatest discrepancy being less than a $5\%$. We present the resulting approximations for holographic Einstein gravity, free scalars and fermions. For the last two, we perform lattice calculations of   $F_{(e)}$ for $e=\sqrt{3}/2\simeq 0.866 $, $e=2 \sqrt{2}/3\simeq 0.943$, $e= \sqrt{15}/4\simeq 0.968$ and $e=2\sqrt{6}/5 \simeq 0.98$ corresponding to values for which the approximations are worse in the case of the EMI model and compare them with our trial function predictions, finding good agreement. 

\subsection{General-CFT approximations for arbitrary eccentricity }
Ellipses of semi-major and semi-minor axes $a$,$b$ and eccentricity $e\equiv \sqrt{1-b^2/a^2}\in [0,1)$ can be parametrized in polar and Cartesian coordinates respectively by
\begin{equation}
r(\theta)=\frac{b}{\sqrt{1-e^2\cos^2\theta}}\, , \quad \text{and } \quad [x(t),y(t)]=[a \cos(t), b \sin(t)]\, ,
\end{equation}
where the ``eccentric anomaly'' $t$ is related to the polar angle $\theta$ by $\tan (\theta) = \tfrac{b}{a} \tan (t)$.
As we vary $e$, we can interpolate between the regime which is very close to the disk (as $e\rightarrow 0$) and the one which approaches the strip-like behavior described in subsection \ref{stripi} (as $e\rightarrow 1$). In the former case, one finds, from Mezei's formula
\begin{equation}\label{eli0}
F_{(e)}  \overset{e\rightarrow 0}{\simeq} F_0 +e^4 \frac{ \pi^4 C_{\ssc T}}{64  }{}+\mathcal{O}(e^8)\, .
\end{equation}
In the opposite regime, we can take advantage of \req{strip}, namely, of the fact that very thin regions are controlled by the strip coefficient $k^{(3)}$ times the ratio of their long and thin dimensions. Using the equation for the width of the squashed ellipse, $2y=2b\sqrt{1-x^2/a^2}$, we can then approximate $F_{(e \rightarrow 1)} $ as \cite{Allais:2014ata}
\begin{equation} \label{elie1}
F_{(e)}  \overset{e\rightarrow 1}{\simeq} k^{(3)} \int_{-a}^{a} \frac{\diff x}{2 b \sqrt{1- x^2/a^2}}=\frac{ k^{(3)}  \pi}{2 \sqrt{1-e^2}} = \frac{ k^{(3)}  \pi}{2\sqrt{2} \sqrt{1-e}} + \mathcal{O}(\sqrt{1-e}) \, .
\end{equation}
This is valid for general CFTs and, at least for some models, it in fact provides a very good approximation to the exact result for not-so-squashed ellipses ---see comments below \req{kemi}.  
Similar formulas valid for general CFTs can be analogously obtained for other closed squashed entangling regions.

%In the previous subsection we considered the regime $e\sim1$. In the opposite limit, \ie for $e\sim 0$, we can use Mezei's formula to obtain 
%When $e$ is small, we can use Mezei's formula to obtain 
%Naturally, for small eccentricities 
Formulas \req{elie1} and \req{eli0} provide us with approximations to $F_{(e)} $ in those two regimes, for general CFTs provided we know the coefficients $k^{(3)}$, $F_0$ and $C_{\ssc T}$. 
Using them, we can try to do better and produce expressions which approximate $F_{(e)} $ for arbitrary values of the eccentricity. 

Our strategy is to build an approximation from a linear combination of test functions with particular relative coefficients. The idea is to fix those coefficients in a way such that   \req{elie1} and \req{eli0} hold when the full expression is expanded around $e\rightarrow 0$ and $e\rightarrow 1$ respectively.\footnote{Let us mention that a similar strategy was used in \cite{Bueno3} (see also \cite{Helmes:2016fcp}) in order to produce high-precision approximations to the entanglement entropy corner functions of general CFTs using the parameters $C_{\ssc T}$ and $k^{(3)}$. In that case, the available (numerical) results for free fields and holography allowed the authors to verify the excellent agreement between the trial functions and the exact curves. } In the former case, this implies setting to zero the $\mathcal{O}(e^1,e^2,e^3,e^5,e^6,e^7)$ coefficients. Naturally, there are many possible candidate functions one may consider, but we find it convenient to choose as building blocks functions of the form $\sim E[e^2]^k/\sqrt{1-e^2}$ with $k$ some positive integer and others $\sim (1-e^2)^{l/2}$ with $l=-1,1,3,\dots$ %We choose
%\begin{equation}
%F_{{(e)}}|^{\rm trial} = \sum_{i} A_i f_i(e)\, , 
%\end{equation}
%where $A_i$ are constants to be fixed and
%\begin{align}
%f_1(e)&\equiv \frac{E[e^2]}{\sqrt{1-e^2}} \, , & &
%f_2(e)\equiv \frac{E[e^2]^2}{\sqrt{1-e^2}} \, , & &
%f_3(e)\equiv \frac{E[e^2]^3}{\sqrt{1-e^2}}\, ,\\
%f_4(e)&\equiv \frac{E[e^2]^4}{\sqrt{1-e^2}} \, , & &
%f_5(e)\equiv \frac{E[e^2]^5}{\sqrt{1-e^2}} \, , & &
%f_6(e)\equiv \frac{E[e^2]^6}{\sqrt{1-e^2}}\, .%f_4(e)&\equiv  E[e^2] (1-e^2)^{3/2}\, , \quad
%f_5(e)\equiv   E[e^2] (1-e^2)^{5/2}\, , \quad
%f_6(e)\equiv   E[e^2] (1-e^2)^{7/2}\, ,\\
%f_4(e)&\equiv (1-e^2)^{3/2}\, , & & f_5(e)\equiv (1-e^2)^{5/2}\, , & & f_6(e)\equiv (1-e^2)^{7/2}\, .
%\end{align}
Functions of these types are such that the $\mathcal{O}(e^1,e^3,e^5,e^7)$ coefficients around $e=0$ are automatically vanishing. For a general linear combination of functions,  expanding around $e=0$ and imposing \req{eli0} fixes four of the coefficients. Expanding around $e=1$ and imposing \req{elie1} fixes an additional one.  These restrictions considerably constrain the possible behavior of the trial functions ---especially if one requires them to be monotonically increasing functions of the eccentricity--- but still allow for some room of variance for intermediate values of $e$. We construct the following two trial functions,
  %to hold at leading order in $(e-1)$ or impose also the subleading term with the logarithm. In the first case, we only require five functions, so we simply set $A_6=0$. In the second, we also fix the logarithmic piece to take the form appearing in \req{elie1}. We get then two trial functions which are given respectively, by 
% whereas doing the same around $e=1$ and imposing \req{elie1} fixes the other two. The final result reads
\begin{equation}\label{trialf}
F_{{(e)}}|^{{\rm tri}_{1} } = \frac{ \alpha_1(e) F_0 +  \alpha_2(e)  C_{\ssc T}+  \alpha_3(e)k^{(3)}}{\alpha_4(e)} \, , \quad F_{{(e)}}|^{{\rm tri}_{2} } = \frac{\beta_0+ \beta_1(e) F_0 +  \beta_2(e)  C_{\ssc T}+  \beta_3(e)k^{(3)}}{\beta_4(e)}\, ,
%\frac{E[e^2] \left[(1-e^2)^2 (64+112 e^2+151e^4) F_0+16\pi^2 e^6(4-3e^2)k^{(3)}+C_{\ssc T}\pi^4 e^4(1-e^2)^2 \right]
%}{32\pi \sqrt{1-e^2}}\, .
\end{equation}
where\footnote{$E[x]$ is the complete elliptic integral, $E[x]\equiv \int_0^{\pi/2} \sqrt{1-x \sin^2 \theta} \diff \theta$. }
\begin{align}
\alpha_1(e)\equiv &+ 2 \Big[ \pi^3 \left(-120+\pi [204+\pi (17\pi-114)]\right) \\ \notag & \quad  \quad  - \pi^2 [-688+\pi \left(1088-504\pi +17\pi^3\right)] E[e^2]\\ \notag & \quad  \quad  +6\pi \left(-176+208\pi-84\pi^3+19\pi^4\right) E[e^2]^2 \\   \notag & \quad  \quad +2 \left(288-2\pi^2 [312+17\pi(3\pi-16)]\right)E[e^2]^3 \\  \notag &\quad  \quad +8 \left(-72+\pi [132+\pi (15\pi-86)] \right)E[e^2]^4\Big]\, , \\ 
\alpha_2(e)\equiv & -\pi^4 (\pi-2)^2  \left( \pi -2 E[e^2] \right)^2 \left( E[e^2]-1\right)  \left( \pi (\pi-6)+(20-6\pi) E[e^2]\right) \, , \\ 
\alpha_3(e)\equiv &+\pi^5  \left( \pi -2 E[e^2] \right)^4 \, , \\ 
\alpha_4(e)\equiv &+ \frac{2\pi^4 (\pi- 2)^4 \sqrt{1-e^2}}{E[e^2]} \, ,
\end{align}
and
\begin{align}
\beta_0(e)\equiv & +\frac{2}{5}\Big[ \pi^2 \left( -8 (32-48e^2+e^4)+\pi^3(17-18e^2+e^4) \right. \\  \notag  & \quad  \quad \left. -2\pi^2 (96-80e^2+7e^4)+4\pi (128-112 e^2+9e^4) \right)  \\  \notag &  \quad  \quad+2 \pi(-1024+128\pi+23\pi^3)E[e^2]-4(-1280+64\pi^2+25\pi^3)E[e^2]^2 \\  \notag  & \quad  \quad -8(640-256\pi+15\pi^2)E[e^2]^3\Big] \, , \\
\beta_1(e)\equiv & +2\pi^2 \left(48-184\pi+84\pi^2-9\pi^3 + e^2 (-24+\pi (44-42\pi+9\pi^2))\right)\\  \notag &  +4E[e^2](7\pi^2(32-3\pi^2)+2E[e^2] \left[-64+7\pi^2(5\pi-16)+(64-6\pi^2)E[e^2] \right])\, , \\
%+4  \left( E[e^2]-1\right)^2 \Big[ \pi^3 (120-228\pi+138\pi^2-11\pi^3)  \\ \notag & \quad  \quad +2 \pi^2 (-344+692\pi-458\pi^2+51 \pi^3)E[e^2]   \\ \notag & \quad  \quad  - 4\pi (-264+572\pi-422\pi^2+53\pi^3)E[e^2]^2  \\ \notag & \quad  \quad + 8 (-72+156\pi -126\pi^2 +17\pi^3)E[e^2]^3    \Big] \, ,  \\ 
\beta_2(e)\equiv & +\pi^6 \left(48-128\pi + 52\pi^2-5\pi^3+(\pi-2)^2 (5\pi-14)e^2\right) \\  \notag &  \quad  \quad +2\pi^4E[e^2] \left[ \pi (128+40\pi-9\pi^3)+2E[e^2](-176+\pi^2(13\pi-20) \right. \\ \notag & \quad  \quad  \left.+2(88+(\pi-32)\pi) E[e^2]) \right] \, , \\
%+2\pi^4(\pi-2)^2\left(\pi-2E[e^2] \right)^2  \left( E[e^2]-1\right)^2 \left(\pi (6+\pi)+2 (\pi-10)E[e^2] \right)   \, , \\ 
\beta_3(e)\equiv & 2\pi^6 (2e^2-1)-28\pi^5 E[e^2]+72\pi^4 E[e^2]^2-16\pi^3E[e^2]^3  \, , \\
%+\pi^5  \left( \pi -2 E[e^2] \right)^4 \left(\pi-10+8E[e^2] \right) \, , \\ 
\beta_4(e)\equiv &4\pi^2(\pi^3-14\pi^2+36\pi-8)\sqrt{1-e^2} \, .& 
%+ \frac{2\pi^4 (\pi- 2)^5 \sqrt{1-e^2}}{E[e^2]} \, .
\end{align}

\begin{figure}[t] \hspace{-0.25cm}
	\includegraphics[scale=0.645]{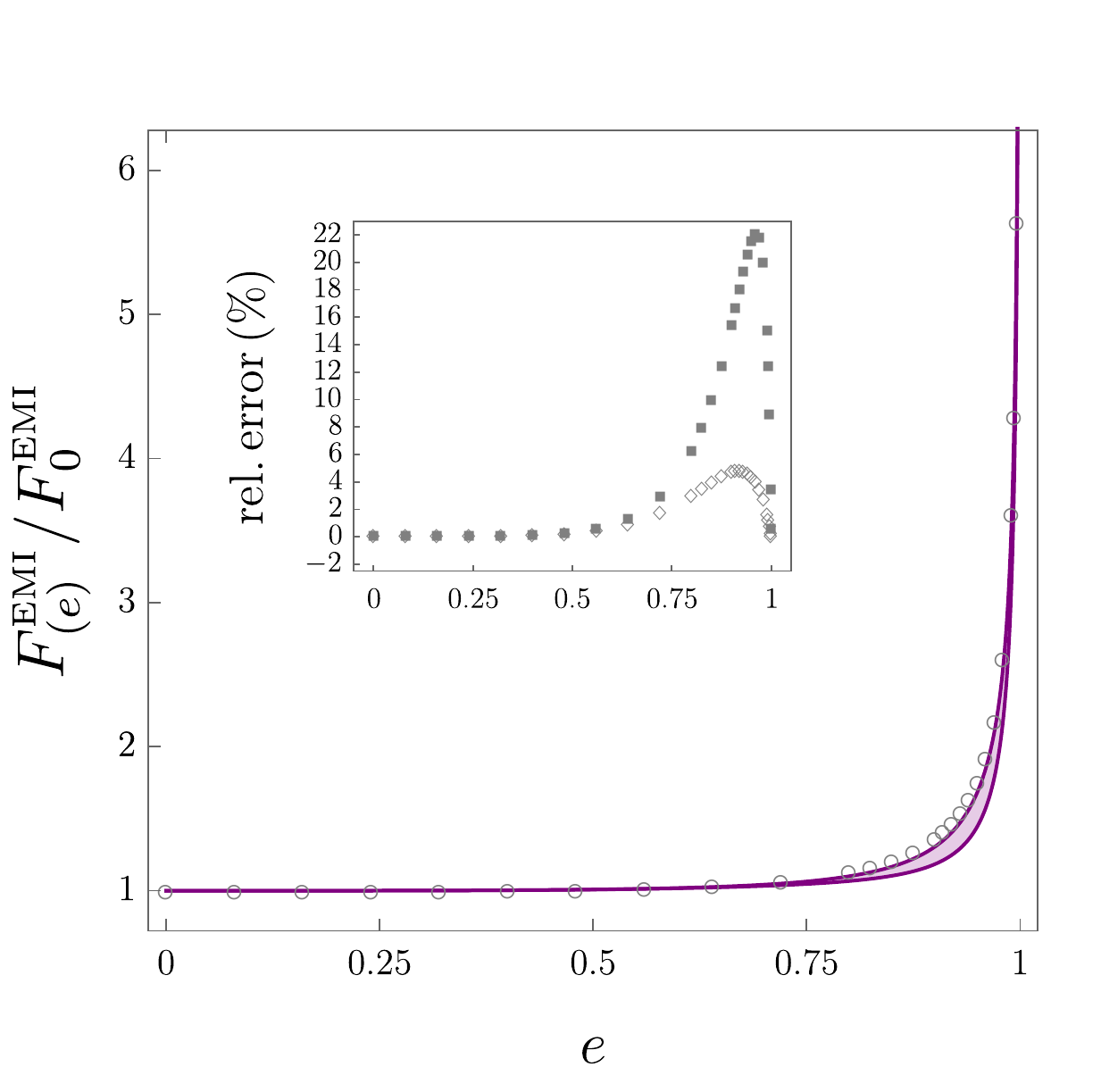}\hspace{-0.2cm}
	\includegraphics[scale=0.64]{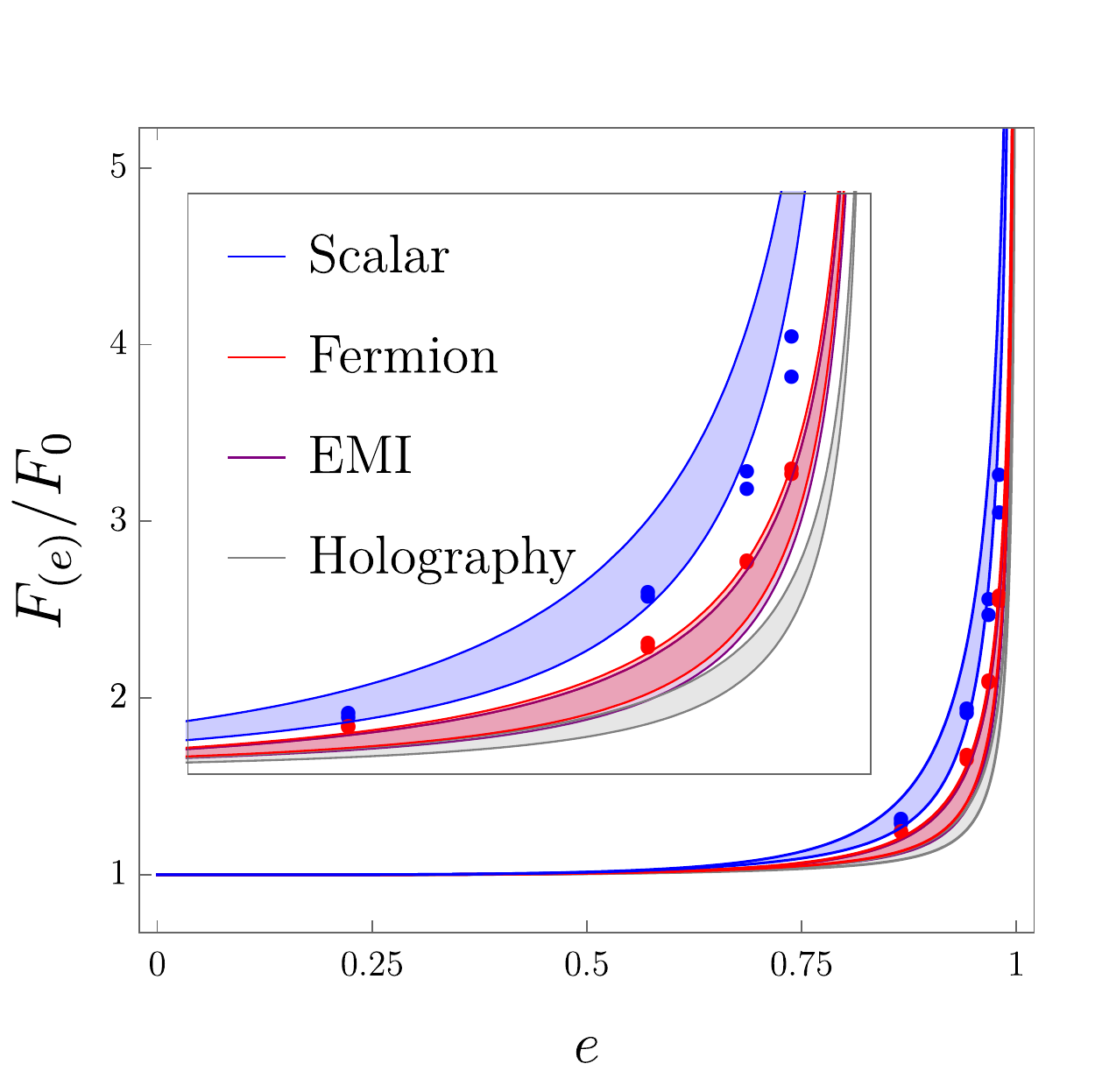}
	\caption{ 	\textsf{We plot the universal contribution to the entanglement entropy corresponding to an elliptic region, normalized by the disk result, $F_0$, as a function of the eccentricity. In the left plot, the data points are (exact) numerical results obtained for the EMI model and the purple curves are the trial functions $F_{(e)}|^{{\rm trial}_{1,2}} $ particularized to this model. The relative error in the approximations is shown in the inset. In the right plot, we present both trial-function curves for: a free scalar, a free fermion, the EMI model and holographic Einstein gravity. The dots correspond to lattice values for the scalar and the fermion. \rd{The inset is a zoom in of the region $e > 0.85$.}}}  %(Lower row) We plot elliptic entangling regions characterized by $f(\theta)=a$, $g(\theta)=b$ for $b=1$ and $a=1,3/2,2,5/2,3,7/2$.   }
	\label{refiss254}
\end{figure}

%\begin{align}\notag
%g_1(e)\equiv& +16 \pi (1-e^2)^2\Big[ 7776-8\pi(240+89 \pi)+5e^4(844+3\pi(9\pi-128))\\ \notag &+4e^2(1796+\pi(13\pi-720))\Big]+ 573440 \left[E[e^2]-1\right]^2E[e^2] \, ,\\ \notag
%g_2(e)\equiv & -2\pi^5(1-e^2)^2  \Big[384 (\pi-2)^2+96e^2 (\pi-2)(5\pi-14)\\ \notag &+e^4 (556+\pi(123\pi-544)) \Big]+6144 \pi^4  \left[E[e^2]-1\right]^2E[e^2] \, ,\\ \notag
%g_3(e)\equiv &+(1-e^2)^2\pi^4 \Big[128(83\pi-183)+16 e^2(359\pi -956)+e^4(2775\pi -7676) \Big] \\  \notag &+ 64\pi^2 E[e^2] \Big[ \pi(471\pi-2480)-3212 [E[e^2]-2]E[e^2]\Big] \, ,\\
%g_4(e)\equiv & +128 \pi [3212 + \pi (471\pi-2480)]\sqrt{1-e^2}   \, .
%\end{align}
Although $F_{(e)}|^{{\rm tri}_{1,2}} $ do not look particularly simple, note that the exact $F_{(e)}$ for a given CFT will in general be an extremely complicated function of $e$ depending on many details of the theory. Our approximations above are completely explicit and depend on the CFT under consideration exclusively through the three constants $k^{(3)}$, $F_0$ and $C_{\ssc T}$. Anticipating results presented in Section \ref{secemi}, we can test the precision of our formulas against numerical calculations for the EMI model. For this, we know $k^{(3)}$, $F_0$ and $C_{\ssc T}$ analytically ---see \req{kemi}, \req{F0emi} and \req{CTEMI} below--- and we can also compute $F_{(e)}$ exactly (up to numerical precision). The first function, $F_{(e)}|^{{\rm trial}_{1}}$ produces a very good approximation to the exact EMI model results. As shown in the inset of the left plot in Fig.\,\ref{refiss254}, the greatest discrepancy is lower than $\sim 5\%$ and it is much smaller for most values of $e$.
The case of $F_{(e)}|^{{\rm trial}_{2}}$ is rather different. For that one, the discrepancy grows up to a maximum of $\sim 22\%$ for $e\sim 0.96$. In view of this, we expect $F_{(e)}|^{{\rm trial}_{1}}$ to be a better approximation for general CFTs. However, we decided to keep $F_{(e)}|^{{\rm trial}_{2}}$ as it does a much better job in approximating the free scalar results obtained in the lattice, as we explain below. %As shown in Fig.\,\ref{refiss254}, the combination of both functions defines a region of possible values of $F_{(e)}$ for a given $e$  

 %However, we expect them to produce precise approximations to the exact curves for general theories. 

%The comparison is shown in the left plot of Fig. \ref{refiss254}. The agreement between the trial function $2$ and the exact data is excellent: for all values of the eccentricity, the discrepancy is lower than $\sim 1 \%$. On the other hand, the agreement with the trial function $1$ is a bit worse for certain values of $e$, and reaches a maximum of almost $\sim 5\%$.
As anticipated, there are other CFTs for which we know the values of the three relevant coefficients and for which  we can therefore compute $F_{(e)}|^{{\rm tri}_{1,2}} $. In particular, for a free scalar \cite{Casini:2009sr,Osborn:1993cr,Klebanov:2011gs,Marino:2011nm}, a free fermion \cite{Casini:2009sr,Osborn:1993cr,Klebanov:2011gs,Marino:2011nm} and Einstein gravity holography \cite{Ryu:2006ef,Liu:1998bu} we have
\begin{align}\label{ctf0s}
&k^{(3)}_{\rm ferm}\simeq 0.0722 \, , & & C_{\ssc T}^{\rm ferm}=\frac{3}{16\pi^2} \, , & & F_0^{\rm ferm}=\frac{1}{8}\left[ 2\log 2+\frac{3}{\pi^2}\zeta(3)\right] \, , \\ \label{ctf0s1}
&k^{(3)}_{\rm scal}\simeq 0.0397 \, ,& & C_{\ssc T}^{\rm scal}=\frac{3}{32\pi^2} \, , & &F_0^{\rm scal}=\frac{1}{16} \left[2\log 2 -\frac{3}{\pi^2} \zeta(3)\right] \, , \\
&k^{(3)}_{\rm holo}=\frac{\pi \Gamma[\tfrac{3}{4}]^2}{\Gamma[\tfrac{1}{4}]^2}\frac{L^2_\star}{G} \, , & & C_{\ssc T}^{\rm holo}=\frac{3}{\pi^3}\frac{L^2_\star}{G} \, ,& & F_0^{\rm holo}=\frac{\pi L^2_\star}{2G} \, . 
\end{align}
Note that the fermion results correspond to a Dirac field, so the values of the coefficients per degree of freedom would require dividing them by $2$.
Using these coefficients, in the right plot of Fig.\,\ref{refiss254} we present the corresponding $F_{(e)}|^{{\rm tri}_{1,2}}  $ curves, alongside the EMI one. As we can see, the fermion and EMI curves are very close to each other\footnote{This is not surprising. For more on the relation between the free fermion and the EMI model see \cite{Agon:2021zvp}.} and, for each model, both trial functions are also rather similar. The scalar and holography curves are the ones making $F/F_0$ greater and lower, respectively, for general values of $e$. This hierarchy of theories is similar to the one encountered for the entanglement entropy corner function $a(\theta)$ normalized by $C_{\ssc T}$ in \cite{Bueno1}. In all cases, $F_{(e)}|^{{\rm tri}_{2}}$ lies notably below $F_{(e)}|^{{\rm tri}_{1}} $ in the whole range. 

% Note also that the case of the scalar is a bit special in the sense that $F_{(e)}|^{{\rm tri}_{2}}$ lies notably above $F_{(e)}|^{{\rm tri}_{1}}  $ for values of $e$ between 0.75 and 0.99, approximately. This has to do with the fact that this model has the greatest quotient $k^{(3)}/F_0$ amongst the theories considered and the choice of including (or not) the subleading piece in \req{elie1} has a greater impact in the approximations. 

%Giving the level of agreement between the exact EMI curve and its corresponding trial functions, we are confident the curves plotted for the different theories should similarly produce good-precision approximations to the exact ones. L%For the ellipses case considered here, we are not aware of additional previous  results for any of these theories (or others) that we can compare with, which in fact makes our approximations more valuable. 

 In the present case, we  have also performed lattice calculations for the free scalar and the fermion corresponding to $e=\sqrt{3}/2\simeq 0.866$, $e=2\sqrt{2}/3\simeq 0.943 $, $e \sqrt{15}/4\simeq 0.968$ and $e=2\sqrt{6}/5 \simeq 0.98$. As explained in the following subsection, we obtain two values of $F_{(e)}$ for each eccentricity and model. They are both presented in Fig.\,\ref{refiss254}. %The results agree reasonably well with the trial functions. 
 We find
\begin{align} \notag
\bar F_{(\sqrt{3}/2)}^{\rm ferm}|_{\rm \ssc lattice}&\simeq  \{1.25 ,  1.25 \}  \, ,  \quad && \bar F_{(\sqrt{3}/2)}^{\rm scal}|_{\rm \ssc lattice}\simeq  \{1.32,  1.30 \}   \, . \\ \notag
\bar F_{(2\sqrt{2}/3)}^{\rm ferm}|_{\rm \ssc lattice}&\simeq \{1.66,  1.69 \}  \, ,  \quad && \bar F_{(2\sqrt{2}/3)}^{\rm scal}|_{\rm \ssc lattice}\simeq    \{1.95,  1.92 \}  \, , \\ \notag
\bar F_{(\sqrt{15}/4)}^{\rm ferm}|_{\rm \ssc lattice}&\simeq \{2.11,  2.10 \} \, ,  \quad && \bar F_{(\sqrt{15}/4)}^{\rm scal}|_{\rm \ssc lattice}\simeq   \{2.48,  2.57 \}   \, . \\ \label{elips}
\bar F_{(2\sqrt{6}/5)}^{\rm ferm}|_{\rm \ssc lattice}&\simeq  \{2.59,  2.56 \} \, ,  \quad && \bar F_{(2\sqrt{6}/5)}^{\rm scal}|_{\rm \ssc lattice}\simeq  \{3.06,  3.27 \}  \, ,
\end{align}
where we defined $\bar F_{(e)} \equiv  F_{(e)}/F_0$ here to avoid the clutter. Now, for the trial functions we find
 \begin{align} 
\bar F_{(\sqrt{3}/2)}^{\rm ferm}|^{{\rm  tri}_1}&  \simeq  1.21  \, ,  \notag
\quad   \bar F_{(2\sqrt{2}/3)}^{\rm ferm}|^{{\rm  tri}_1}\simeq 1.62  \, , \quad  \bar F_{(\sqrt{15}/4)}^{\rm ferm}|^{{\rm  tri}_1}\simeq 2.10\, , \quad \bar F_{(2\sqrt{6}/5)}^{\rm ferm}|^{{\rm  tri}_1}\simeq 2.60\, , \\ \notag
\bar F_{(\sqrt{3}/2)}^{\rm ferm}|^{{\rm  tri}_2}&  \simeq  1.13  \, ,  
\quad   \bar F_{(2\sqrt{2}/3)}^{\rm ferm}|^{{\rm  tri}_2}\simeq 1.41  \, , \quad  \bar F_{(\sqrt{15}/4)}^{\rm ferm}|^{{\rm  tri}_2}\simeq 1.80\, , \quad \bar F_{(2\sqrt{6}/5)}^{\rm ferm}|^{{\rm  tri}_2}\simeq 2.25\, , \\ \notag
\bar F_{(\sqrt{3}/2)}^{\rm scal}|^{{\rm  tri}_1}&  \simeq 1.43   \, ,  \notag
\quad   \bar F_{(2\sqrt{2}/3)}^{\rm scal}|^{{\rm  tri}_1}\simeq 2.31  \, , \quad  \bar F_{(\sqrt{15}/4)}^{\rm scal}|^{{\rm  tri}_1}\simeq 3.30 \, , \quad \bar F_{(2\sqrt{6}/5)}^{\rm scal}|^{{\rm  tri}_1}\simeq 4.32 \, , \\
\bar F_{(\sqrt{3}/2)}^{\rm scal}|^{{\rm  tri}_2}&  \simeq 1.27  \, ,  
\quad   \bar F_{(2\sqrt{2}/3)}^{\rm scal}|^{{\rm  tri}_2}\simeq 1.87   \, , \quad  \bar F_{(\sqrt{15}/4)}^{\rm scal}|^{{\rm  tri}_2}\simeq 2.67 \, , \quad \bar F_{(2\sqrt{6}/5)}^{\rm scal}|^{{\rm  tri}_2}\simeq 3.57 \, .
%F_{(\sqrt{3}/2)}^{\rm ferm}|^{{\rm  tri}_2}&\simeq 1.26  F_0^{\rm ferm} \, ,  
%&& F_{(2\sqrt{2}/3)}^{\rm ferm}|^{{\rm  tri}_2}\simeq 1.72  F_0^{\rm ferm} \, , \\
%F_{(\sqrt{3}/2)}^{\rm scal}|^{{\rm  tri}_1}&\simeq 1.43  F_0^{\rm scal} \, ,  
%&& F_{(2\sqrt{2}/3)}^{\rm scal}|^{{\rm  tri}_1}\simeq 2.31  F_0^{\rm scal} \, , \\
%F_{(\sqrt{3}/2)}^{\rm scal}|^{{\rm  tri}_2}&\simeq 1.72 F_0^{\rm scal} \, ,  
%&& F_{(2\sqrt{2}/3)}^{\rm scal}|^{{\rm  tri}_2}\simeq 2.87  F_0^{\rm scal} \, .
\end{align}
We observe that the fermion results never differ from the first trial function prediction more than a $\sim 4\%$ and in most cases the agreement is better. The agreement with the second trial function is much worse, and the discrepancies reach a maximum of $\sim 17\%$. In the case of the scalar, it is the second trial function the one which approximates  better the lattice results, with disagreements lower than $\sim 5\%$ in most cases, except for the last value, which differs by a $\sim 13\%$. On the other hand, the results are clearly off from the ones predicted by the first trial function: in some cases, the discrepancy grows up to $\sim 40\%$.
We expect the results both for the scalar and the fermion to have an uncertainty which is probably not less than $\sim 5\%$. This does not explain however why we cannot fit the results for both theories with a single curve. In fact, we suspect that the lattice is considerably underestimating the actual scalar results. In particular, as we mention below, analogous calculations for the disk region yield $\bar F_0^{\rm ferm}|_{\rm \ssc lattice}\simeq 0.99$ and $\bar F_0^{\rm scal}|_{\rm \ssc lattice}\simeq 0.92$. Namely, while the fermion result approaches the analytical answer very well, the scalar one is an $8\%$ lower. In view of this and of the agreement for the EMI and free fermion results with the first trial function, a reasonable guess is that $F_{{(e)}}|^{{\rm tri}_{1} }$ is actually a very good approximation to the exact curve for general CFTs, including the scalar, whereas  $F_{{(e)}}|^{{\rm tri}_{2} }$ is not so much. We say a bit more about this in the next subsection.

%\comment{HEEEREEEEEE}
%In the fermion case, the second trial function provides a better approximation. The lattice results differ from this one by $\sim 3\%$ and $\sim 6\% $, whereas the difference grows to  $\sim 7\%$ and $\sim 11\% $ when we compare with the first one. The situation is reversed for the scalar. In that case, the results differ from the first one by $\sim 1.6\%$ and $\sim 1.7\% $ whereas the difference grows to $\sim 15\%$ and $\sim 18\%$ when we compare with the second trial function. As we explain later, the lattice results have an uncertainty which is probably not less than $\sim 5 \%$, so the results seem to be as good as they can be. %\comment{more blah blah}
%Besides, we expect our approximations $F_{(e)}|^{{\rm tri}_{1,2}} $ to yield precise approximations to the actual curves for general theories.

\subsection{Free-field calculations in the lattice}
As anticipated, in this subsection we compute $F_{(e)}/F_0$ for elliptic regions in the lattice for free scalars and fermions. As argued in \cite{Casini:2015woa}, trying to obtain $F_0$ from a direct calculation of the entanglement entropy in a square lattice does not produce reasonable results.\footnote{For satisfactory calculations in radial lattices see \cite{Liu:2012eea,Klebanov:2012va}.} Rather, one obtains wildly oscillating answers as the ratio $R/\delta$ varies (here $\delta$ stands for the lattice spacing).  The problem has to do with the fact that we cannot resolve the radius of the disk with a precision better than $\delta$, which means that we cannot distinguish disks with radii $R$ and $R( 1+ a \delta)$, with $a \sim \mathcal{O}(1)$. But such uncertainty will pollute $F$ via the area-law piece in the entanglement entropy \req{entro} as $-F \rightarrow -F + 2\pi c_0 a$. As it is clear from this, the issue cannot be resolved by making the disk radius larger in the lattice.

\begin{figure}[t]  \centering
	\includegraphics[scale=0.525]{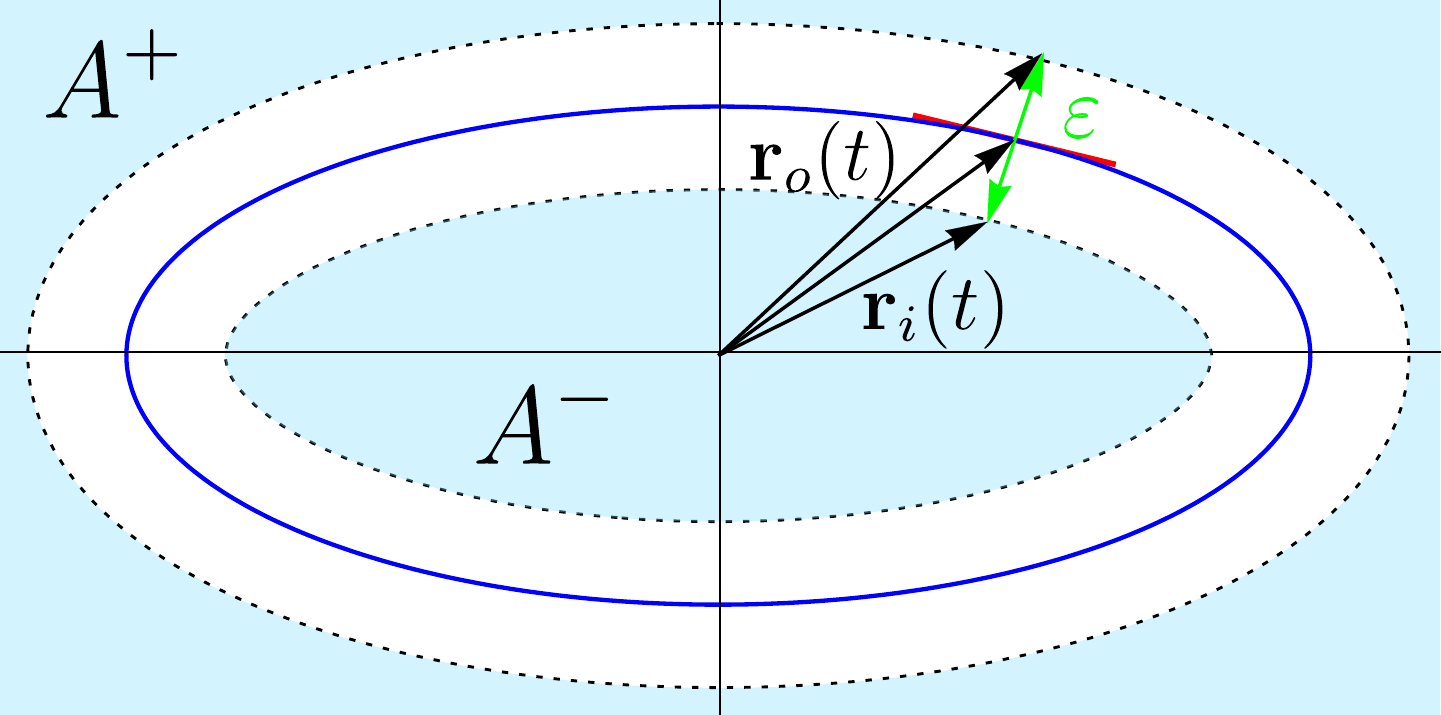}
		\includegraphics[scale=0.605]{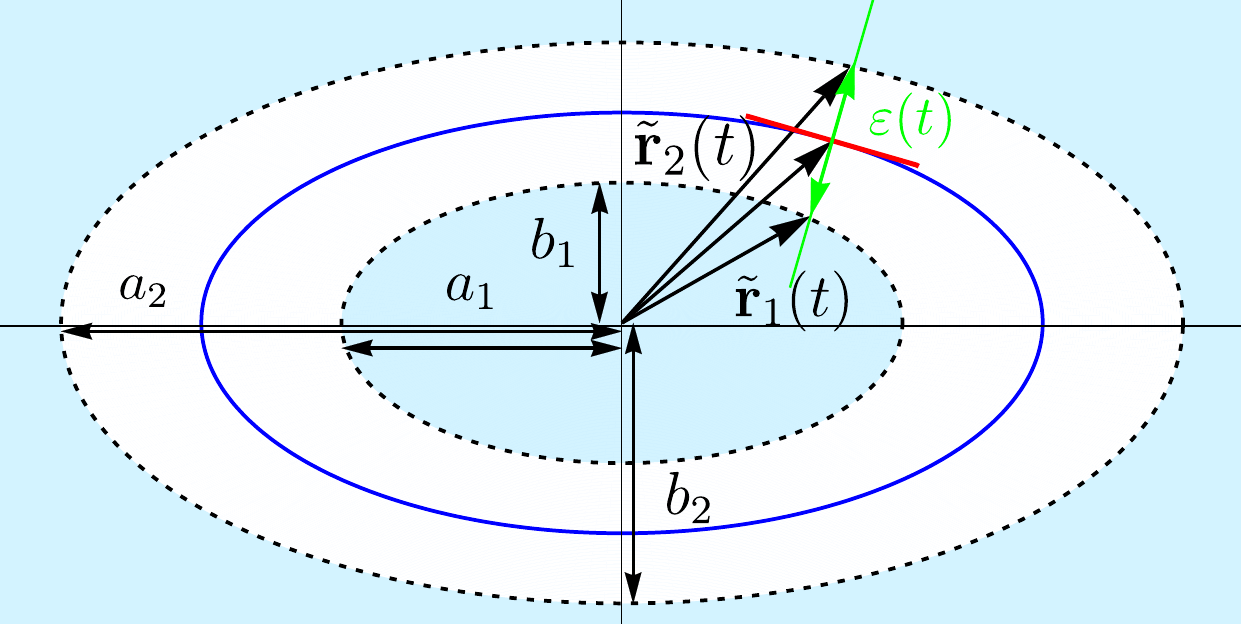}
	\caption{ \textsf{(Left) We plot an example of the geometric setup required for the evaluation of the EE for an elliptic region $A$ (solid blue curve) using the MI of an inner region $A^-$ and the complement of one which contains it, $\overline{A^+}$, as in \req{annulii}. In this case, the inner and outer regions are defined so that $\varepsilon$ is constant for all values of $t$. The points ${\bf r}_i(t)$ and ${\bf r}_o(t)$ are determined by moving a distance $\varepsilon/2$ inwards and outwards respectively from each point of $\partial A$.  (Right) We show an alternative setup for which $\overline{A^+}$ and $A^-$ are both elliptic regions, which forces $\varepsilon(t)$ to be variable. The points $\tilde {\bf r}_{1,2}(t)$ are determined from the intersection of the normal to $\partial A$ at a given point with the inner and outer ones.}  }
	\label{refiss34}
\end{figure}

An alternative approach which does produce convergent results consists in using mutual information as a geometric regulator \cite{Casini:2007dk,Casini:2008wt,Casini:2014yca,Casini:2015woa}. Given a region $A$, we consider two concentric ones $A^-$ and $\overline{A^+}$. These are defined by considering a normal to $\partial A$ at each point and moving a distance $\varepsilon/2$ inwards and outwards along such direction, respectively. $\varepsilon$ can be chosen to be constant for all values of the parameter $s$ defining $\partial A$ or, alternatively, it can be a function of $s$ if one decides that $A^+$ and $\overline{A^-}$ should be dilated versions of $A$ ---\eg if $A$ is an ellipse, $A^-$ and $\overline{A^+}$ would also be ellipses using this second method.  In both cases, we are left with an inner annulus of width $\varepsilon$ (constant or variable) ---see Fig.\,\ref{refiss34}.

%--
% such that the former is the complement of a slightly dilated version of the latter, with an inner annulus of width $\varepsilon$ separating both ---see Fig.\,\ref{refiss34}. 

The idea is then to consider the MI
\begin{equation}
I(A^+,A^-)=\see(A^+)+\see(A^-)-\see(A^+ \cup A^-)\, ,
\end{equation}
which, using the purity of the ground state can be rewritten as
\begin{equation}\label{annulii}
I(A^+,A^-)=\see(\overline{A^+})+\see(A^-)-\see( \overline{A^+ \cup A^-})\, ,
\end{equation}
where %$\overline{A^+}$ is now the dilated version of $A^-$, and 
$\overline{A^+ \cup A^-}$ is the annulus region. The three regions appearing in \req{annulii} are finite and are thus defined by a finite number of points in the lattice, so they are suitable for that setup.

 Now the idea is to consider annuli such that $\varepsilon/R\ll 1$ while keeping $\varepsilon/\delta \gg 1$. In that limit, the MI behaves as $I(A^+,A^-) \simeq 2 \see (A)$, where $A$ is the intermediate region. This equality is true up to terms which diverge as $\sim 1/\varepsilon$ and $\sim 1/\delta$ respectively in the corresponding limits. Hence, in order to extract $F$ from the MI (\ref{annulii}), one must deal with those terms first. As argued in \cite{Casini:2015woa}, if we parametrize the boundary  of $A$ by the length parameter $s$, one finds for  $\varepsilon/R\ll 1$,
 \begin{equation}\label{ik3}
 I(A^+,A^-)=k^{(3)} \int_{\partial A} \frac{\diff s}{\varepsilon (s)}-2 F + \text{subleading in } \varepsilon\, , 
 \end{equation} 
where $k^{(3)}$ is the coefficient characterizing the EE of a thin strip region ---see \req{strip}. In order for this to work, the region $A$ must be chosen precisely half way between $\partial A^+$ and $\partial A^-$. As mentioned earlier, $\varepsilon (s)$ corresponds to the separation between $\partial A^+$ and $\partial A^-$ at a given point $s$, measured along the normal direction to $\partial A$ at that point. Therefore, given some region $A$ in the lattice, we can extract its $F$ by computing $I(A^+,A^-)$ as in \req{annulii} and subtracting the first piece in the r.h.s. of \req{ik3}, then dividing by $-2$. For fixed values of $\varepsilon/R$, where $R$ is some characteristic size of $A$, the results will improve their accuracy as $R/\delta \gg 1$.

In the case of disk regions, \req{ik3} simplifies considerably, and one gets
 \begin{equation}\label{ik23}
 F_0=-\frac{1}{2}\left[  I({\rm disk }_{R_2},{\rm disk}_{R_1})-k^{(3)} \frac{2\pi R}{\varepsilon} \right]\, ,
 \end{equation} 
 where $\varepsilon\equiv R_2-R_1$ and $R\equiv (R_1+R_2)/2$. In this case, both methods described above involve a constant $\varepsilon$ and $A^-$, $\overline{A^+}$ are always disks.
  
 For an elliptic region $A$ of eccentricity $e$, on the other hand, we can either choose a constant $\varepsilon$ or, alternatively, force $A^-$ and $\overline{A^+}$ to be ellipses as well. In both cases, 
computing $\varepsilon$ requires obtaining the line which intersects the boundary of the elliptic entangling region normally at the point determined by $t$ ---see Fig.\,\ref{refiss34}. Assuming the ellipse is centered at the origin of coordinates and parametrized by $[a \cos(t),b \sin(t)]$, this is given by
\begin{equation}
y(x)=\frac{a}{b} \tan(t)  x + \frac{b^2-a^2}{b} \sin(t)\, .
\end{equation}
%where $a= (a_1+a_2)/2$, $b=(b_1+b_2)/2$.
 A point on the normal can be parametrized as $[a \cos(t),y(a \cos(t))]$ and we want to determine the quantity $\alpha$ that we need to add to both coordinates so that the new point lies a distance $\varepsilon/2$ from $[a \cos(t), b \sin(t)]$. Hence, we need to solve
 \begin{equation}
 [a \cos (t) - (a \cos (t)+ \alpha)]^2 + [b \sin (t) - y(a \cos(t)+ \alpha)]^2 =\varepsilon/2\, ,
 \end{equation}
 for $\alpha$. Solving this equation and plugging back we find that the points ${\bf  r}_{i,o}(t)$ defining the shapes of the inner and outer curves read
 \begin{eqnarray} \label{rio}
{\bf  r}_{i,o}(t)=
 \begin{cases}
\left[a \cos(t)+\frac{ (-1)^i  b \varepsilon}{2\sqrt{b^2+a^2\tan^2(t)}},y\left(a \cos(t)+\frac{ (-1)^i  b \varepsilon}{2\sqrt{b^2+a^2\tan^2(t)}}\right)\right]\quad \text{for} \quad \frac{3\pi}{2}<t\leq\frac{\pi}{2} \, ,\\
\left[a \cos(t)-\frac{ (-1)^i  b \varepsilon}{2\sqrt{b^2+a^2\tan^2(t)}},y\left(a \cos(t)-\frac{  (-1)^i  b \varepsilon}{2\sqrt{b^2+a^2\tan^2(t)}}\right)\right]\quad \text{for} \quad\frac{\pi}{2}< t\leq\frac{3\pi}{2}\, .
\end{cases}
\end{eqnarray}
Here, $(-1)^i=-1,1$ for the inner and outer shapes, respectively. Finally, we can compute the entanglement entropy of the original elliptic region as
 \begin{equation}\label{ik130}
 F_{(e)}=-\frac{1}{2}\left[  I({\rm pseudoellipse }_{i},{\rm pseudoellipse}_{o})-k^{(3)} \frac{4 a E[e^2]}{\varepsilon}\right]\, ,
 \end{equation}  
where $I({\rm pseudoellipse }_{i},{\rm pseudoellipse}_{o})$ is the mutual information of the ``pseudoellipses'' defined by the shapes ${\bf  r}_{i}(t)$ and ${\bf  r}_{o}(t)$ respectively.

  %The result reads
% \begin{equation}
% \alpha= \pm \frac{b \varepsilon}{ \sqrt{b^2+a^2 \tan^2(t)}}\, ,
 %\end{equation}
% and from this
 
For the second method, on the other hand, we consider two concentric ellipses of the same eccentricity $e$ parametrized by $[a_{1,2}\cos(t) ,b_{1,2}  \sin(t)]$. Computing $\varepsilon(t)$ requires now obtaining the intersections of the normal line to $\partial A$ with the exterior and interior ellipses. These have equations $y=b_{1,2} \sqrt{1-x^2/a_{1,2}^2}$ where now $a= (a_1+a_2)/2$, $b=(b_1+b_2)/2$. The intersections occur at the points 
% line which intersects the intermediate ellipse normally at a point $t$ ---see Fig.\,\ref{refiss34}. 
%Assuming the ellipses are centered at the origin of coordinates, this is given by
%\begin{equation}
%y=\frac{a}{b} \tan(t)  x + \frac{b^2-a^2}{b} \sin(t)\, ,
%\end{equation}
%where $a= (a_1+a_2)/2$, $b=(b_1+b_2)/2$.
%The relevant intersection of this line with the exterior and interior ellipses, with equations $y=b_{1,2} \sqrt{1-x^2/a_{1,2}^2}$, occur at the points 
${\bf \tilde r}_{1,2}(t)$ ---see Fig.\,\ref{refiss34}--- where
\begin{align} \label{x1}
\tilde x_1&\equiv \frac{a a_1^2(a^2-b^2)\sin(t)\tan(t) + \sqrt{a_1^2 b^2b_1^2\left[b^2b_1^2+a^2a_1^2\tan^2(t)- (a^2-b^2)^2 \sin^2(t)\right]}}{a^2a_1^2\tan^2(t)+b^2b_1^2} \, , \\ \label{x2}
\tilde y_1&\equiv \frac{b^2 b_1^2(b^2-a^2)\sin(t)+a\tan(t)  \sqrt{a_1^2 b^2b_1^2\left[b^2b_1^2+a^2a_1^2\tan^2(t)- (a^2-b^2)^2 \sin^2(t)\right]}}{a^2a_1^2b\tan^2(t)+b^3b_1^2} \, ,
\end{align}
and $(\tilde x_2,\tilde y_2)$ are given by the same expressions replacing $(1\leftrightarrow 2)$  in all labels.
From this, $\varepsilon(t)$ can be obtained as
\begin{equation} \label{epis}
\varepsilon(t)=\sqrt{[\tilde x_1(t)-\tilde x_2(t)]^2+[\tilde y_1(t)-\tilde y_2(t)]^2}\, .
\end{equation}
Putting the pieces together, we find that the EE universal term for an ellipse of eccentricity $e$ with a boundary half way between two concentric ellipses of the same eccentricity can be obtained  as
 \begin{equation}\label{ik13}
 F_{(e)}=-\frac{1}{2}\left[  I({\rm ellipse }_{2},{\rm ellipse}_{1})-k^{(3)} \int_0^{\frac{\pi}{2}} \diff t \frac{2(a_1+a_2)\sqrt{1-e^2 \cos^2 t} }{\varepsilon(t)}\right]\, ,
 \end{equation} 
where $1,2$ are labels referring to the two (inner and outer, respectively) concentric ellipses.

Both \req{ik130} and \req{ik13} should produce equivalent approximations to $F_{{(e)}}$ for sufficiently large values of $b/\varepsilon$. However, as we discuss in Appendix \ref{EMIMIregu} in the case of the EMI model ---for which we can perform calculations for arbitrary values of $b/\varepsilon$--- the prescriptions are not equally good in their precision for finite values of $b/\varepsilon$ ---see Fig.\,\ref{ref44}. In particular, we find that the pseudoellipses one produces more stable and accurate answers, specially for values of $e$ close to $1$. In the lattice, we have evaluated $I({\rm pseudoellipse }_{i},{\rm pseudoellipse}_{o})$ as well as  $I({\rm ellipse }_{2},{\rm ellipse}_{1})$ numerically using \req{annulii} and then subtracted the purely geometric pieces proportional to $k^{(3)}$ in each case. Similarly to the EMI case, we observe that the first method produces greater and more stable results, so we only present those here.   %\comment{blaaaaaah}

We perform  lattice calculations for free scalars and fermions. For the former, we consider  a set of  fields and conjugate momenta $\phi_i, \pi_j$, $i, j= 1,\dots, N$  labelled by their positions at the square lattice. These satisfy   canonical  commutation relations,  $[\phi_i ,\pi_j]=i\delta_{ij}$ and $[\phi_i,\phi_j]=[\pi_i,\pi_j]=0$. Then, given some  Gaussian state   $\rho$, the   EE can be computed from the two-point correlators
%Let us start with the free scalar.  %studying how the tripartite information scales with the ratio of the entangling region characteristic length $R$ and the distance between the regions $r$. 
%Consider a square lattice of $N$ points and a set of scalar fields and momenta $\phi_i,\pi_j$, $i,j=1,\dots,N$ satisfying canonical commutation relations, $[\phi_i,\pi_j]=i\delta_{ij}$, $[\phi_i,\phi_j]=[\pi_i,\pi_j]=0$. Given a Gaussian state $\rho$, consider the two-point correlators
$
X_{ij}\equiv \tr (\rho \phi_i\phi_j)$ and   $P_{ij}\equiv \tr (\rho \pi_i \pi_j)\, .
$
In particular, one  has \cite{2003JPhA...36L.205P,Casini:2009sr}
\begin{equation}\label{see}
\see(A)=\tr \left[(C_A +1/2) \log (C_A    +1/2)- (C_A -1/2)   \log (C_A-1/2) \right]\, , 
\end{equation}
where $C_A \equiv \sqrt{X_A    P_A}$ and  $(X_A)_{ij}\equiv X_{ij}$, $(P_A)_{ij}=P_{ij}$ (with $i,j\in A$) are the restrictions of the correlators   to the  sites belonging  to region $A$.

%Then, the entanglement entropy corresponding to a region $A$ can be obtained from the restrictions of $X_{ij}$ and $P_{ij}$ to the sites belonging to such a region as
%\begin{equation}\label{see}
%\see(A)=\tr \left[(C_A+1/2) \log (C_A+1/2)- (C_A-1/2)\log (C_A-1/2) \right]\, , 
%\end{equation}
%where $C_A \equiv \sqrt{X_A P_A}$ and  we denote $(X_A)_{ij}\equiv X_{ij}$, $(P_A)_{ij}=P_{ij}$ with $i,j\in A$.

%Here we will work in $d=2+1$, so each index $i$ corresponds to coordinates in a two-dimensional lattice. 

The lattice  Hamiltonian reads  in this case
\begin{equation}
H= \frac{1}{2}   \sum_{n,m=-\infty}^{\infty}   \left[\pi^2_{n,m}   + (\phi_{n+1,m}   -\phi_{n,m})^2  +(\phi_{n,m+1}-\phi_{n,m})^2 \right]\, ,
\end{equation}
where the lattice spacing   is set to one. The   relevant expressions for the correlators $X_{(x_1,y_1),(x_2,y_2)}$ and $P_{(x_1,y_1),(x_2,y_2)}$   corresponding to the vacuum state are given by \cite{Casini:2009sr} 
\begin{align}
X_{(0,0),(i,j)} & = \frac{1}{8\pi^2}\int_{-\pi}^{\pi} \diff  x \int_{-\pi}^{\pi} \diff y \frac{\cos (jy) \cos (ix) }{\sqrt{2(1-\cos x)+2(1-\cos y)}}\, , \\
P_{(0,0),(i,j)} & = \frac{1}{8\pi^2}\int_{-\pi}^{\pi} \diff  x \int_{-\pi}^{\pi} \diff y  \cos(jy)  \cos(ix) \sqrt{2(1-\cos x)+2(1-\cos y)}\, .
\end{align}
%Using these expressions, we can evaluate the tripartite information of lattice regions $A$, $B$, $C$ using \req{see} and the general expression 
%\begin{align}\notag
%I_3(A,B,C) =& +\see(A)+\see(B)+\see(C) \\ &-\see(AB)-\see(AC)-\see(BC)+\see(ABC)\, . 
%\end{align}

In the case of the free fermion, we consider fields  $\psi_i$, $i=1,\dots, N$  defined at the lattice points  and satisfying canonical  anticommutation   relations, $\{\psi_i, \psi_j^{\dagger} \}=\delta_{ij}$. Given a Gaussian state  $\rho$, we define the matrix of correlators  $D_{ij} \equiv \tr \small( \rho \psi_i \psi_j^{\dagger} \small)$. Then, similarly to the scalars case,  the EE for some region $A$ can be obtained from the restriction of $D_{ij}$ to the corresponding  lattice sites as \cite{Casini:2009sr}
\begin{equation}
\see (A)=- \tr   \left[ D_A   \log D_A   + (1-D_A) \log (1-D_A)\right] \, .
\end{equation}
The lattice Hamiltonian is given   by
%The three-dimensional lattice Hamiltonian we consider for the free fermion reads
\begin{equation}
H=-\frac{i}{2}  \sum_{n ,m} \left[  \left(\psi^{\dagger}_{m, n}   \gamma^0     \gamma^1 (\psi_{m+1,n}  -\psi_{m,n})+\psi^{\dagger}_{m,  n}   \gamma^0 \gamma^2   (\psi_{m,n+1}   -\psi_{ m,n}  ) \right)  - h.c.\right] \, ,
\end{equation}
and the correlators in the vacuum state read \cite{Casini:2009sr} 
\begin{equation}
D_{(n,k),(j,l)} =   \frac{1}{2}\delta_{n,j} \delta_{kl}     - \int_{-\pi}^{\pi} \diff x  \int_{-\pi}^{\pi} \diff y   \frac{\sin (x) \gamma^0 \gamma^1+   \sin(y) \gamma^0 \gamma^2}{8\pi^2 \sqrt{   \sin^2 x +   \sin^2 y}} e^{i(x (n-j)+y(k-l))}\, .
\end{equation}

\begin{figure}[t] \hspace{-0.3cm}
	 \includegraphics[scale=0.64]{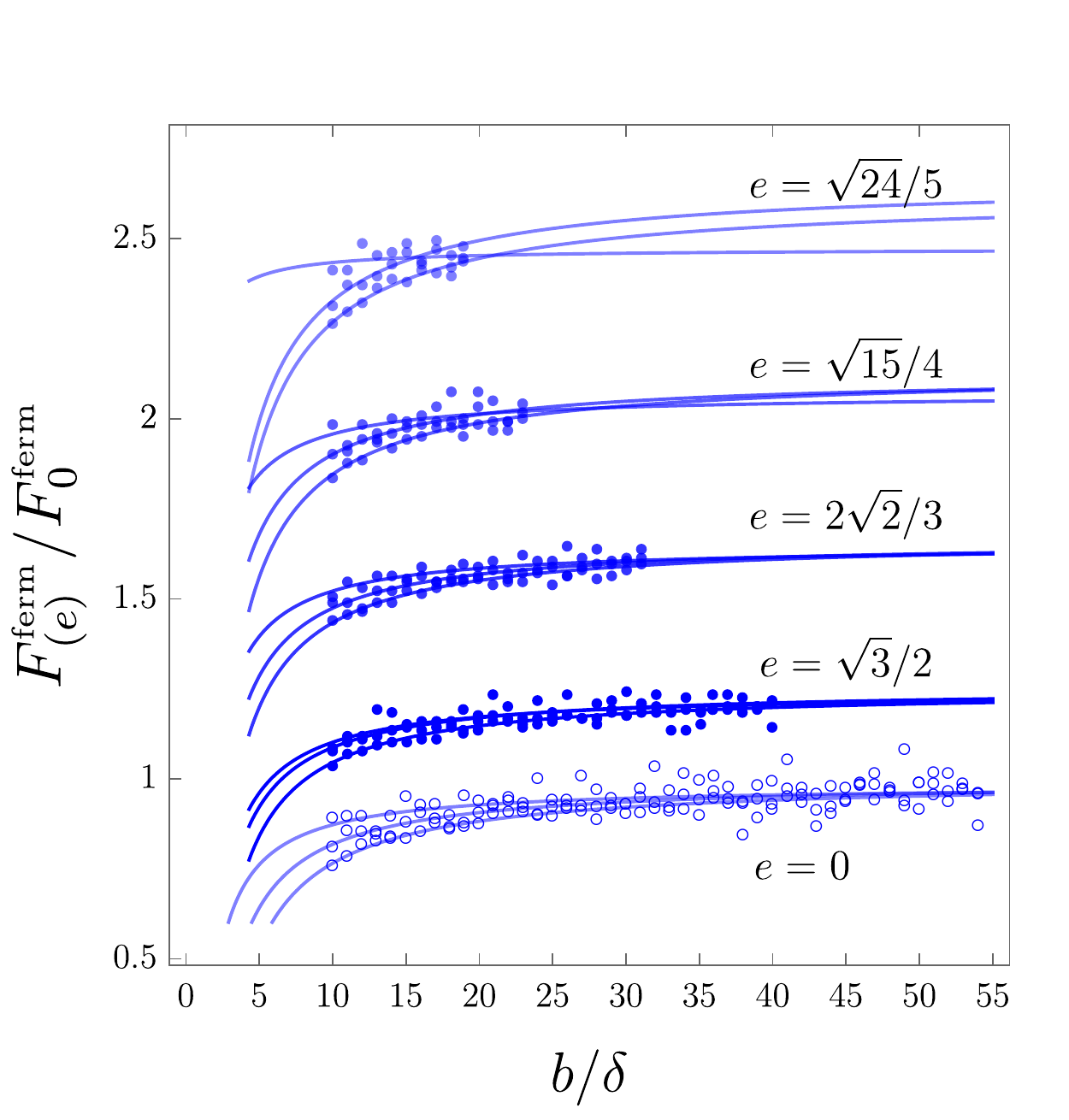}\hspace{-0.2cm}	
	\includegraphics[scale=0.64]{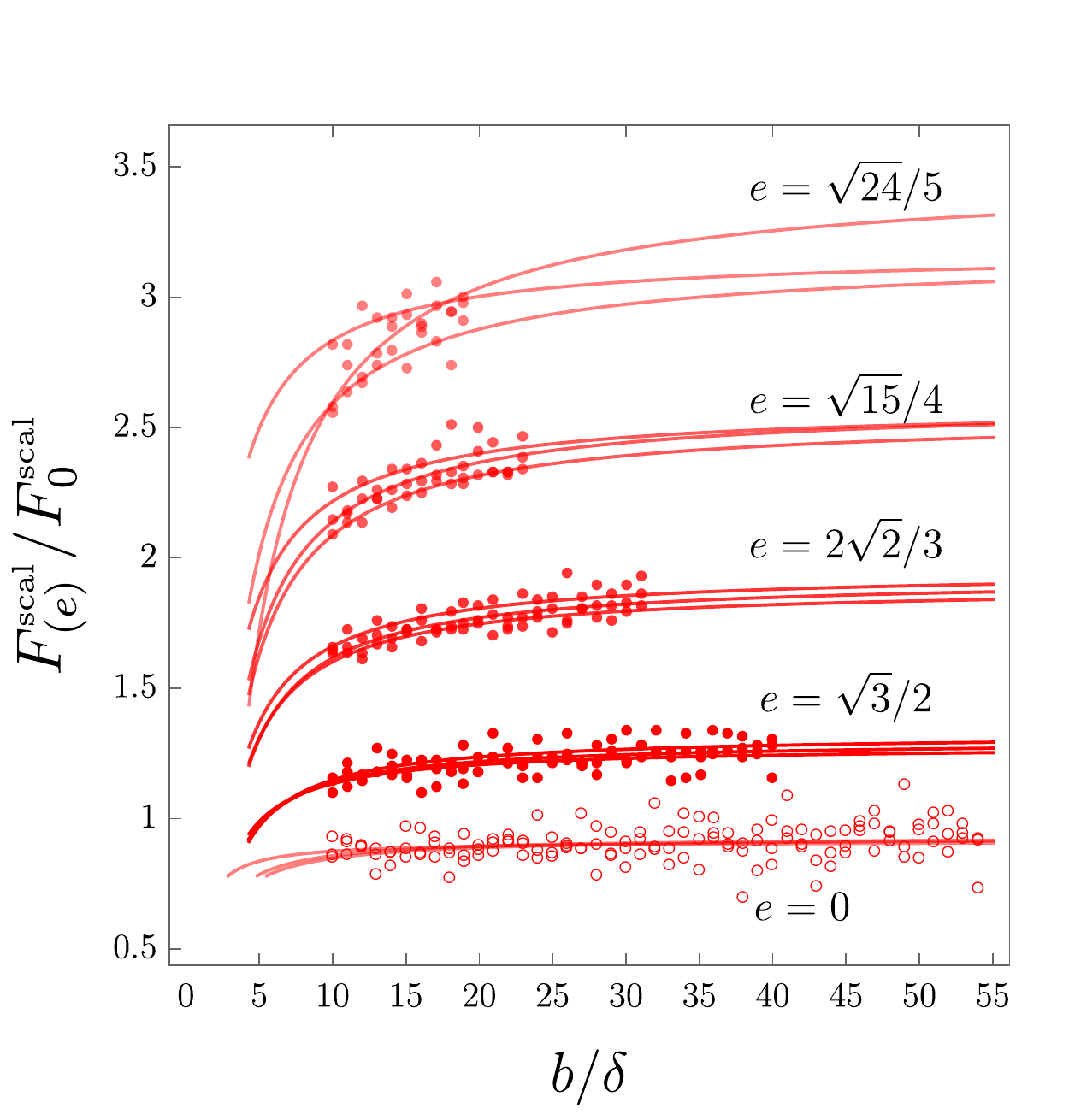}
	\caption{ \textsf{We plot $F_{(e)}$ normalized by the disk result for free fermions and scalars in the lattice for eccentricities $e=0,\sqrt{3}/2\simeq 0.943 ,2\sqrt{2}/3\simeq0.968 ,\sqrt{15}/4\simeq 0.968, 2\sqrt{6}/5\simeq 0.98$, as a function of the ellipses' semiminor axes divided by the lattice spacing. The data points are obtained using the constant-$\varepsilon$ formula \req{ik130} for $\varepsilon=6,8,10$. The solid lines correspond to $\{1/x,1\}$ fits  performed ---for each value of $e$--- to the data points for fixed $\varepsilon$. } %The dotted lines correspond to the trial functions defined in \req{trialf} for each eccentricity. 
	 }
	\label{refiss344}
\end{figure}

As a warm up, we perform calculations for disk regions ---see also \cite{Casini:2015woa}, where this was previously done. In that case, the results for $F_0$ are known analytically both for the scalar and the fermion ---see \req{ctf0s} and \req{ctf0s1} above. In the lattice, we consider pairs of disks of different radii and the corresponding annuli and use \req{ik23} to extract our results. The results have a numerical uncertainty associated to the finiteness of both $\varepsilon/R$ and $\delta/\varepsilon$. This comes from the difficulty of achieving $\varepsilon/R \ll 1$ while keeping $\delta/\varepsilon \ll 1$. Naturally, 
 considering sufficiently large values of $\varepsilon$ requires taking disks with large enough radii, which increases the computation time. In Fig.\,\ref{refiss344} we present various data points of $F_0$ (normalized by the analytic result) for both the scalar and the fermion as a function of $b/\varepsilon$. We take $\varepsilon/\delta=6,8,10$ and values of disk radius up to $R/\delta =58$. As we can see, the results oscillate considerably for both models. In order to extract a number for each of them, we proceed in two ways. First, we take the two values of $F_{(e)}$ corresponding to the largest $R/\delta$ considered for each of the three values of $\varepsilon$, and take the average of those six values. Then, we multiply them by $1$ plus a small correction factor ($\sim 0.05$) which  we obtain in Appendix \ref{EMIMIregu} based on the EMI model and which should approximately account for the fact that we are considering not too large values of $R/\delta$ for the given values of $\varepsilon/\delta$.  Secondly, we perform fits  $\{1/x,1 \}$ to the series of data points for each $\varepsilon$ and take the average of the three values.  We find from these two methods, respectively,
\begin{equation}
F_0^{\rm scal}|_{\rm \ssc lattice}\simeq \{0.98,0.92\} F_0^{\rm scal} \, ,  \quad F_0^{\rm ferm}|_{\rm \ssc lattice}\simeq  \{1.02,0.99\} F_0^{\rm ferm}\, .
\end{equation}
The results are in very good agreement with the expectations, although we observe that the scalar one obtained from the fits is still underestimating the actual result.

%\comment{HEEEREEE}

% for small values of this quotient but tend to approach the analytic results rather well as  $\varepsilon/\delta$ grows. In order to extract a number from our calculations, we consider the results for the three largest $R_2$ in each case (we consider disks up to $R_2/\delta =63$ both for the scalar and the fermion). From those, we take the three values which minimize the quantity $\sqrt{(\varepsilon/R)^2+(\delta/R)^2}$. Taking the average of those nine values for each model, we find
%\begin{equation}
%F_0^{\rm scal}|_{\rm \ssc lattice}\simeq 1.01 F_0^{\rm scal} \, ,  \quad F_0^{\rm ferm}|_{\rm \ssc lattice}%\simeq  0.98 F_0^{\rm ferm}\, .
%\end{equation}
%The results are therefore rather close to the analytic ones. 

Moving on to the ellipses, we perform calculations for eccentricities $e=\sqrt{3}/2\simeq 0.866$, $e=2\sqrt{2}/3\simeq 0.943$, $e=\sqrt{15}/4\simeq 0.968$ and $e= 2\sqrt{6}/5\simeq 0.98$. These belong to the region in which the almost-round and very-thin approximations ---\req{eli0} and \req{elie1} respectively--- do not work so well. As mentioned earlier, the pseudoellipses method yields better results than the variable-$\varepsilon$ one, so we use the former to perform our calculations. The results are shown in Fig.\,\ref{refiss344}. Proceeding similarly to the disks, in order to extract approximations for  large-$b/\delta$ results we make fits to the data for fixed values of $\varepsilon$ and take the average. Alternatively, we also consider for each   $\varepsilon$ the values obtained for the two greatest values of $b/\delta$, take the average of the six and multiply by the correction factor. The results appear  in \req{elips} above. The fermion data points look somewhat more harmonious than the scalar ones and the data series produce more similar predictions. On the other hand, as mentioned in the previous subsection, our candidate trial function $F_{{(e)}}|^{{\rm tri}_{1} }$ ---which yielded a very good agreement with the exact EMI results--- produces results which also agree very well with the fermion ones in the lattice, but not so for the scalar. It is possible that the methods we are using are still insufficient to account for the underestimation of the results associated to the finite-$R/\delta$ limitations in the case of the free scalar. If that is case, the predictions of $F^{\rm scal}_{{(e)}}|^{{\rm tri}_{1} }$ are probably more accurate than our lattice results. Alternatively, it may also be that the idea of accurately reproducing $F_{(e)}$ for any CFT with a single function completely determined by $k^{3}$, $C_{\ssc T}$ and $F_0$ does not work so well and one rather requires a set of functions ---\eg $F^{\rm scal}_{{(e)}}|^{{\rm tri}_{1} }$, $F^{\rm scal}_{{(e)}}|^{{\rm tri}_{2} }$ and the intermediate curves between them, as displayed in Fig.\,\ref{refiss254}--- to precisely account for all possible theories.

Regarding the fundamental question which is subject of study in the present paper, the evidence  gathered from holography \cite{Fonda:2014cca}, the EMI model, the lattice results for free fields and the two limits ($e \rightarrow 0$ and $e \rightarrow 1$) for general theories makes it clear that $F_{(e)}$ is  a monotonically increasing function of the eccentricity for a given CFT ---and therefore $F_{(e)}\geq F_0$ for all $e$ for general CFTs.  As a final comment for this section, we also point out that the methods utilized here for computing EE using MI should be similarly applicable to other regions beyond ellipses. In that case, the constant-$\varepsilon$ method appears to be the best choice.

%In particular, our trial functions \req{trialf} display their greatest errors for the EMI model for the range $e \in (0.8,0.98)$, approximately ---see inset in the left plot of Fig.\,\ref{refiss254}. In Fig.\,\ref{refiss344}  we present the lattice results for different values of $(b_2-b_1)/\delta$ alongside the predictions coming from our trial functions. As mentioned earlier, even though the lattice calculations seem to converge reasonably well, they 

%have an uncertainty which comes from the difficulty of achieving $\varepsilon/R \ll 1$ while keeping $\delta/\varepsilon \ll 1$. In our calculations, we have reached simultaneous  values of $\varepsilon/R, \delta/\varepsilon\sim 0.12$, so it is reasonable to expect that the averages performed in order to obtain the values presented in \req{elipf} and \req{elips} can well have uncertainties of $\sim 5$-$10\%$. This is compatible with the discrepancies observed with the trial functions discussed above.

%In order to do this, we make use of mutual information as a geometric regulator.

 %the raw results obtained from a direct extraction of the constant term computation of the entanglement entropy of disk regions in a square lattice for free fields wildly oscillate 

\section{More general shapes in the EMI model}\label{secemi}
In order to obtain explicit results for more complicated entangling regions, we consider now the so-called 
 ``Extensive mutual information model'' (EMI) \cite{Casini:2008wt}. This follows from considering a general formula for the mutual information compatible with the known general axioms satisfied by this measure in a general QFT ---see \eg \cite{Agon:2021zvp}--- plus the additional requirement that it is an extensive function of its arguments, namely, $I(A,B)+I(A,C)=I(A,BC)$ for any $A,B,C$. This model corresponds to a free fermion in $d=2$ \cite{casini2005entanglement} but does not describe the mutual information of any actual theory (or limit of theories) for $d\geq 3$, as recently shown \cite{Agon:2021zvp}. Nonetheless, the fact that it satisfies all known properties expected for a valid MI ---in particular, some highly non trivial, such as monotonicity: $B \subseteq C  \Rightarrow I(A,B) \leq I(A,C)$--- makes it a useful toy model which has shown to produce qualitatively and quantitatively reasonable results in various cases \cite{Casini:2015woa,Bueno1,Witczak-Krempa:2016jhc,Bueno:2019mex,Estienne:2021lxh}.     % which satisfies the standard axioms 
 %As its name suggests, this is defined through a formula for the mutual information which is extensive in the entangling regions (so that $I_3(A,B,C)\equiv 0$ for all $A,B,C$)

In the EMI model,  the entanglement entropy of a  region $A$ in a time slice of $3$-dimensional Minkowski spacetime is given by
\begin{equation} \label{emiee}
\SEMI (A)=\kappa_{(3)}\int_{\partial A}\diff\mathbf{r}_1\int_{\partial A}\diff\mathbf{r}_2\, \frac{\mathbf{n}_1\cdot\mathbf{n}_2}{\abs{\mathbf{r}_1-\mathbf{r}_2}^{2}}\, ,
\end{equation}
where the integrals are both along the entangling surface $\partial A$, $\mathbf{n}$ is a unit normal vector and $\kappa_{(3)}$ is a positive parameter. %The usual structure of divergences in the entanglement entropy arises from the contact point $\mathbf{r}_1 \rightarrow \mathbf{r}_2$.

There are different ways to regulate the above integrals, which otherwise diverge as $\mathbf{r}_1 \rightarrow \mathbf{r}_2$. For our purposes, it will be convenient to introduce a UV regulator $\delta$ along an auxiliary extra dimension, so that we replace  $\abs{\mathbf{r}_1-\mathbf{r}_2}^{2} \rightarrow \abs{\mathbf{r}_1-\mathbf{r}_2}^{2}+\delta^2$ with $\delta \ll $ all the rest of scales.
%\begin{equation}\label{regu}
%\SEMI (A)=\kappa_{(3)}\int_{\partial A}\diff\mathbf{r}_1\int_{\partial A}\diff\mathbf{r}_2\, \frac{\mathbf{n}_1\cdot\mathbf{n}_2}{\abs{\mathbf{r}_1-\mathbf{r}_2}^{2}+\delta^2}\, .
%\end{equation}

Now, in order to obtain a computationally useful formula for the universal constant term, $F^{\rm \ssc EMI}(A)$, note from \req{entro} that, on general grounds,
\begin{equation}\label{delF}
 F(A)+ \mathcal{O}(\delta ) =- \left[\delta \frac{\partial \see (A)}{\partial \delta} + \see(A) \right] \, ,
\end{equation}
\ie the combination in the r.h.s. is equivalent to $F(A)$ up to terms which vanish as the regulator is taken to zero. Alternatively, we can also subtract the area-law piece, fixing first the value of $c_0$ by computing the EE in a particular example. Then, we trivially have
\begin{equation}\label{F2}
 F(A)+ \mathcal{O}(\delta ) =c_0 \frac{L(\partial A)}{\delta} -\see(A) \, .
\end{equation}
Observe that while $c_0$ is regulator-dependent, it does not change as long as we perform all our calculations with the same regulator.

Using \req{delF} we find
\begin{equation}\label{femi}
F^{\rm \ssc EMI}(A)+ \mathcal{O}(\delta )=-\kappa_{(3)}\int_{\partial A}\diff\mathbf{r}_1\int_{\partial A}\diff\mathbf{r}_2\,  \frac{ [| \mathbf{r}_1-\mathbf{r}_2|^2-\delta^2]}{[\abs{\mathbf{r}_1-\mathbf{r}_2}^{2}+\delta^2]^2} \mathbf{n}_1\cdot\mathbf{n}_2\, .
\end{equation}
Alternatively, we can use \req{F2}. In order to fix $c_0$, we can evaluate the EE for a simple region like a strip, or a disk. By doing so, we find that $c_0=\pi$ for the regulator introduced above, and we can write\footnote{One could try to extract a $\delta$-independent expression for $F^{\rm \ssc EMI}(A)$, \eg by rewriting \req{femi} as
\begin{equation}\label{femi0}
F^{\rm \ssc EMI}(A)+ \mathcal{O}(\delta )= -\frac{ \kappa_{(3)}}{2}\int_{\partial A}\diff\mathbf{r}_1\int_{\partial A}\diff\mathbf{r}_2\, \frac{\partial^2}{\partial \delta^2} \log \left[ |\mathbf{r}_1 - \mathbf{r}_2|^2 + \delta^2\right]  \mathbf{n}_1\cdot\mathbf{n}_2\, ,
\end{equation}
trading the derivatives with the integrals and finally taking the $\delta \rightarrow 0$ limit. Unfortunately, we have not been able to obtain anything too useful from this approach. }
\begin{equation}\label{femi2}
F^{\rm \ssc EMI}(A)+ \mathcal{O}(\delta )=\kappa_{(3)} \left[ \frac{\pi}{\delta}\int_{\partial A}\diff\mathbf{r}_1 -\int_{\partial A}\diff\mathbf{r}_1\int_{\partial A}\diff\mathbf{r}_2\, \frac{\mathbf{n}_1\cdot\mathbf{n}_2}{\abs{\mathbf{r}_1-\mathbf{r}_2}^{2}+\delta^2}\right]\, .
\end{equation}

Now, in order to make progress, let us choose a particular set of coordinates. We parametrize our entangling surfaces as 
\begin{align}
\mathbf{r}_i=&f(\theta_i)  \left(\cos\theta_i,  \sin\theta_i\right)\, , \quad i=1,2\, ,
%\mathbf{r}_2=&f_2\left(\cos\theta_2,\sin\theta_2\right),
\end{align}
%\begin{align}
%\mathbf{r}_i=&\left(f(\theta_i) \cos\theta_i, g(\theta_i) \sin\theta_i\right)\, , \quad i=1,2\, ,
%\mathbf{r}_2=&f_2\left(\cos\theta_2,\sin\theta_2\right),
%\end{align}
where $f(\theta_i)$ is a function of the polar angle, $\theta \in [0,2\pi)$.
%and $g(\theta_i)$ are functions of the polar angle, $\theta \in [0,2\pi)$. 
Naturally, in the case of a disk, $f(\theta_i)=R$, where $R$ is just its radius. 
In general we have
%The denominator of EMI for $d=3$ reads
%\begin{align}
%\diff \mathbf{r}_i&=\diff\theta_i \sqrt{(f_i\sin \theta_i -f'_i \cos \theta_i )^2+(g_i\cos \theta_i +g'_i \sin \theta_i)^2} \, , \\
%\abs{\mathbf{r}_1-\mathbf{r}_2}^{2}&=(f_1 \cos \theta_1-f_2 \cos \theta_2)^2+ (g_1 \sin�\theta_1 -g_2\sin %\theta_2)^2\, ,
%\end{align}
%where we used the notation $f_i \equiv f(\theta_i)$, $g_i \equiv g(\theta_i)$. These expressions simplify considerably when the two functions coincide, namely
%Then, we have
%The denominator of EMI for $d=3$ reads
\begin{equation}
\diff \mathbf{r}_i=\diff\theta_i \sqrt{f_i^2+f_i'^2}\quad \text{and}\quad \abs{\mathbf{r}_1-\mathbf{r}_2}^{2}=f_1^2+f_2^2-2f_1f_2\cos\left[\theta_1-\theta_2\right]\, ,
\end{equation}
where we used the notation $f_i \equiv f(\theta_i)$.
As for the normal vectors, one finds
%The normal vectors, defined by $\mathbf{n}=\frac{\partial^2_{\theta_1}\mathbf{r}}{\abs{\partial^2_{\theta_1}\mathbf{r}}}=\frac{\mathbf{r}''}{\abs{\mathbf{r''}}}$ are given by
\begin{equation}
\mathbf{n}_i=-\left(\frac{f_i \cos \theta + f_i' \sin\theta}{ \sqrt{f_i^2+f_i'^2}},\frac{f_i \sin \theta- f'_i \cos \theta}{ \sqrt{f_i^2+f_i'^2}}\right)\, .
%\mathbf{n}_2=&\left(\frac{\left(f''_2-f_2\right)\cos\theta_2-2\sin\theta_2}{\sqrt{\left({f''_2}-f_2\right)^2-f_2'^2}},\frac{\left(f''_2-f_2\right)\sin\theta_2-2\cos\theta_2}{\sqrt{\left({f''_2}-f_2\right)^2-f_2'^2}}\right)
\end{equation}
%where
%\begin{equation}
%|n_i\equiv  \sqrt{f_i^2+f_i'^2} \, .
%\sqrt{(\left(f''_i-f_i\right)\cos\theta_i-2f_i' \sin\theta_i )^2 + (\left(g''_i-g_i\right)\sin\theta_i+2 g_i' \cos\theta_i)^2}\, .
%\end{equation}
%For a single function, this simplifies to
%\begin{equation}
%n_i=\sqrt{\left({f''_i}-f_i\right)^2+4f_i'^2}\, .
%\end{equation}
Using these we get
\begin{equation}
\diff \mathbf{r}_1 \diff \mathbf{r}_2\,  \mathbf{n}_1 \cdot \mathbf{n}_2= \diff \theta_1 \diff \theta_2 \left[\cos (\theta_1-\theta_2) \left[f_1 f_2+f'_1 f'_2\right]+ \sin (\theta_1-\theta_2) \left[f_1' f_2-f_2' f_1\right] \right]\, .
\end{equation}
Plugging the above expressions back in \req{femi} or \req{femi2}, we obtain an explicit formula for the universal term, up to $\mathcal{O}(\delta)$ terms.%, which is now a functional $F^{\rm \ssc EMI}=F^{\rm \ssc EMI}[f(\theta_1),f(\theta_2),\theta_1,\theta_2]$. % which, for simplicity we will denote $F^{\rm \ssc EMI}[f]$ from now on.

%\subsection{Disk }
Let us first consider the simplest possible case, corresponding to a disk region. Then, we have $f_i=R$, and the expression for $F^{\rm \ssc EMI}$ simplifies drastically,
\begin{equation}
F^{\rm \ssc EMI}_0+\mathcal{O}(\delta)=-\kappa_{(3)} R^2 \int_0^{2\pi} \diff \theta_1  \int_0^{2\pi} \diff \theta_2   \frac{ [2R^2[1-\cos(\theta_1-\theta_2)]-\delta^2] \cos(\theta_1-\theta_2)}{[2R^2[1-\cos(\theta_1-\theta_2)]+\delta^2]^2}+\mathcal{O}(\delta) \, .
\end{equation}
The integrals can be explicitly performed and the result reads
\begin{equation}\label{F0emi}
F^{\rm \ssc EMI}_0=2\pi^2 \kappa_{(3)}\, .
\end{equation}
Alternatively, we can use \req{femi2}. The first integral is trivially $2\pi R$ and, for the second, we can use the symmetry of the disk to fix $\mathbf{r}_2=(R,0)$ $\mathbf{n}_2=(1,0)$. Then, we have
\begin{align}
F^{\rm \ssc EMI}_0+\mathcal{O}(\delta)&=\kappa_{(3)} \left[\frac{\pi}{\delta } 2\pi R- 2\pi R^2 \int_0^{2\pi} \diff \theta_1  \frac{\cos \theta_1}{ 4 R^2\sin^2(\theta_1/2)+\delta^2} \right]+\mathcal{O}(\delta) \, ,\\
&=\kappa_{(3)} \left[ \frac{2\pi^2 R}{\delta } - \frac{2\pi^2 R}{\delta }+ 2\pi^2 +\mathcal{O}(\delta)\right]= 2\pi^2 \kappa_{(3)}+\mathcal{O}(\delta)\, ,
\end{align}
finding the same answer.

Now, in order to consider more complicated figures, we will numerically integrate \req{femi} and \req{femi2} for various values of $\delta \ll 1$ and obtain a linear fit to the resulting data ---which is indeed approximately linear in $\delta$. The values of $F^{\rm \ssc EMI}$ are in each case obtained as the $\delta=0$ limits of the fits. 

\subsection{Slightly deformed disks}
The next-to-simplest case corresponds to small deformations of the disk region like the ones considered in \req{fiy}. In that case, we can compare our numerical results to Mezei's formula \req{fmeze0}. 
%For a slightly deformed disk with equation
%\begin{equation}\label{fi}
%f(\theta_i)=R \left[1+ \epsilon \sum_{\ell} \left( \frac{a^{(c)}_{\ell}}{\sqrt{\pi}} \cos (\ell \theta_i) + \frac{a^{(s)}_{\ell}}{\sqrt{\pi}} \sin (\ell \theta_i) \right) \right]\, , \quad \epsilon \ll 1\, ,
%\end{equation} 
%the result for $F$ at leading order in $\epsilon$ is given, for general CFTs in three dimensions by
%\begin{equation}\label{fmeze}
%F=F_0+ \epsilon^2 \, \frac{\pi^3 C_{\ssc T} }{24} \sum_{\ell} \ell (\ell^2-1)
%\left[ (a^{(c)}_{\ell})^2+(a^{(s)}_{\ell})^2 \right] + \mathcal{O}(\epsilon^4) \, , \end{equation}
%where $C_{\ssc T}$ is the stress-tensor two-point function charge.
 In the case of the EMI model, the stress-tensor two-point function coefficient takes the value \cite{Agon:2021zvp}
\begin{equation}\label{CTEMI}
C^{\rm \ssc EMI}_{\ssc T}=\frac{16 \kappa_{(3)}}{\pi^2}\, .
\end{equation}
This can be obtained by considering an entangling region with a straight corner which, for general CFTs, contains a universal term of the form \cite{Casini:2006hu,Hirata:2006jx}
\begin{equation}
S_{\rm \ssc EE}|_{\rm \ssc univ}= - a(\theta) \log (\ell /\delta)\, , 
\end{equation}
where the function of the opening angle $a(\theta)$ satisfies  \cite{Bueno1}
\begin{equation}
a(\theta)=\frac{\pi^2}{24}C_{\ssc T} (\theta-\pi)^2+ \dots 
\end{equation}
for almost smooth corners. 
In the EMI case, an explicit calculation produces \cite{Casini:2008wt,Swingle:2010jz}
\begin{equation}\label{}
a_{\rm \ssc EMI}(\theta)= 2 \kappa_{(3)} [1 +(\pi-\theta) \cot \theta]\, ,
\end{equation}
from which \req{CTEMI} easily follows. Combining this with \req{fmeze0} we have
\begin{equation}\label{FEMI} 
F^{\rm \ssc EMI}=2\pi^2 \kappa_{(3)}+  \epsilon^2 \, \frac{2 \pi \kappa_{(3)}}{3} \sum_{\ell} \ell (\ell^2-1)
\left[ (a^{(c)}_{\ell})^2+(a^{(s)}_{\ell})^2 \right] +\mathcal{O}(\epsilon^4)\, .
 \end{equation}
In order to test the validity of this formula, we start by inserting \req{fiy} in \req{femi}. Even though we have not succeeded in performing the integrals analytically in full generaly, we manage to do so in a case-by-case basis for individual values of $\ell$. We start by expanding around $\epsilon=0$ inside the integrand and then take the limit $\delta \rightarrow 0$ of the $\mathcal{O}(\epsilon)$, $\mathcal{O}(\epsilon^2)$ and $\mathcal{O}(\epsilon^3)$ terms, which turn out to be finite. The resulting expressions can be integrated analytically, and we find that the  $\mathcal{O}(\epsilon)$ and  $\mathcal{O}(\epsilon^3)$ pieces vanish, as expected on general grounds, and an exact match with \req{FEMI} for the quadratic term.
 
  \begin{figure}[t] \centering
	\includegraphics[scale=0.8]{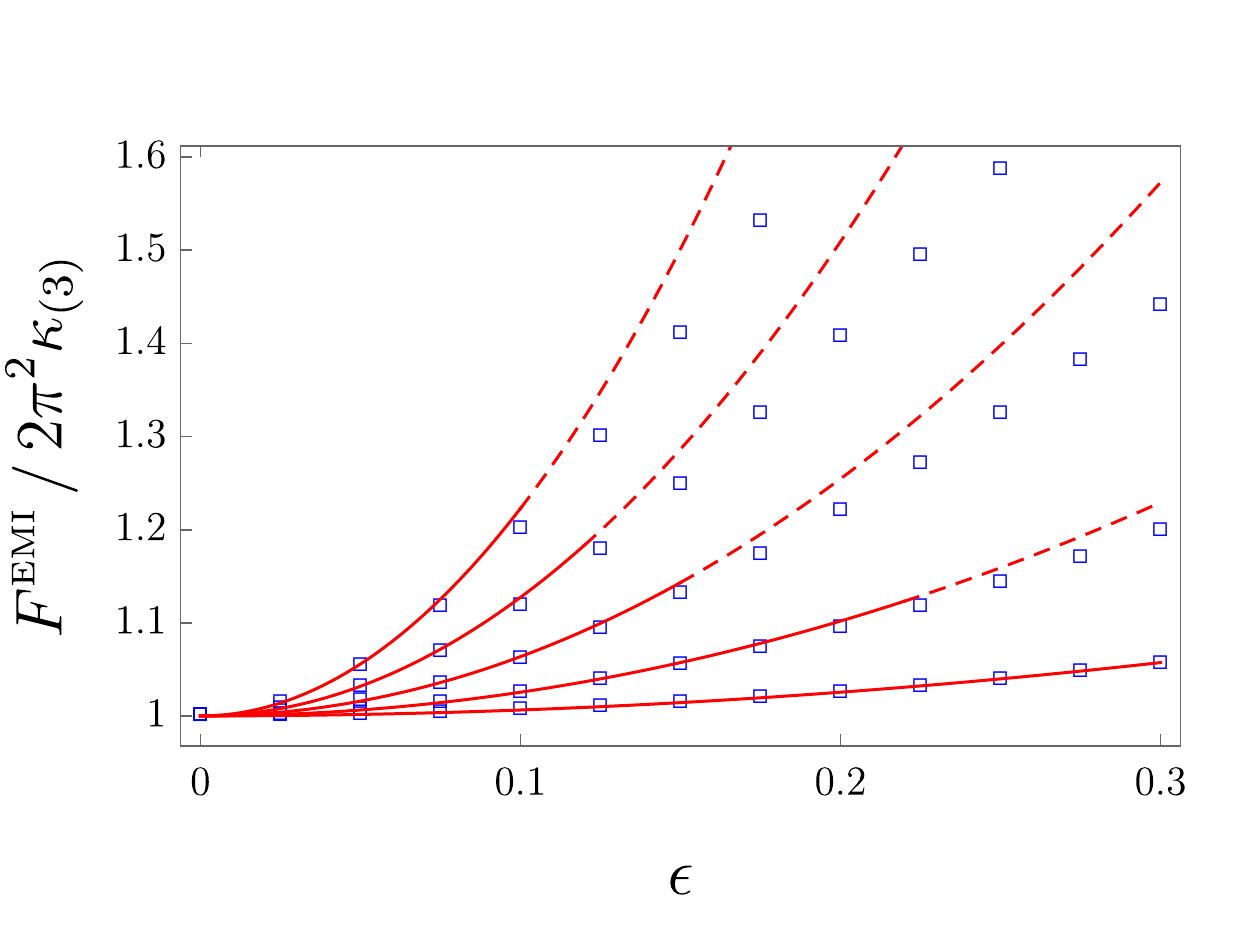}
	\caption{ \textsf{We plot the universal piece $F^{\rm \ssc EMI}$ for small deformations of a disk region corresponding to \req{fami1} for $\ell=2,3,4,5,6$ as a function of  the small parameter $\epsilon$. The blue dots are obtained numerically integrating \req{femi} in each case. The red lines correspond to the leading-order parabolic approximations produced by Mezei's formula \req{FEMI}.}}  %(Lower row) We plot elliptic entangling regions characterized by $f(\theta)=a$, $g(\theta)=b$ for $b=1$ and $a=1,3/2,2,5/2,3,7/2$.   }
	\label{refiss2}
\end{figure}
 
We can also use formula \req{FEMI}  to perform some checks on the numerical methods that we use below for arbitrary (non-perturbative) shapes. In order to extract $F^{\rm \ssc EMI}$ from \req{femi} or \req{femi0} for a given $f(\theta)$, we numerically integrate the corresponding expressions for various small values of $\delta$ and perform a linear fit of the resulting data points. The value of  $F^{\rm \ssc EMI}$ is then obtained as the $y$-intercept of the corresponding line. Doing this for a family of the form 
\begin{equation}\label{fami1}
f_{\ell}(\theta)\equiv R \left[ 1 + \frac{\epsilon}{\sqrt{\pi}} \cos (\ell \theta)\right]\, ,
\end{equation}
 for several values of $\ell$ we obtain results (essentially identical for both \req{femi} and \req{femi0}) which are always compatible with Mezei's formula for small enough values of $\epsilon$, as shown in Fig. \ref{refiss2}. There, we observe that as $\ell$ grows, the quadratic approximation becomes worse. This is not surprising: while the coefficient of the quadratic term grows like $\ell^3$, it is natural to expect higher powers of $\ell$  to appear in the quartic and higher-order pieces. Fitting the data points to expansions involving even powers of $\epsilon$, we can extract the quadratic coefficient and compare with \req{FEMI}. The results are
 \begin{align}
&\bar F^{\rm \ssc EMI}|^{\req{femi}}_{f_{2}(\theta)} \simeq 1 +0.637\epsilon^2+\mathcal{O}(\epsilon^4) \, ,  & & \bar F^{\rm \ssc EMI}|^{\rm \ssc Mezei}_{f_{2}(\theta)}=  1 + \frac{2}{\pi} \epsilon^2 \simeq 1 + 0.637 \epsilon^2 \, ,  \\
&\bar F^{\rm \ssc EMI}|^{\req{femi}}_{f_{3}(\theta)} \simeq 1 +2.55 \epsilon^2 +\mathcal{O}(\epsilon^4)\, , & &  \bar F^{\rm \ssc EMI}|^{\rm \ssc Mezei}_{f_{3}(\theta)}=  1 + \frac{8}{\pi} \epsilon^2 \simeq 1 + 2.55\epsilon^2 \, , \\
&\bar F^{\rm \ssc EMI}|^{\req{femi}}_{f_{4}(\theta)} \simeq 1 +6.35  \epsilon^2 +\mathcal{O}(\epsilon^4) \, ,& &  F^{\rm \ssc EMI}|^{\rm \ssc Mezei}_{f_{4}(\theta)}=  1 + \frac{20}{\pi} \epsilon^2 \simeq 1 + 6.37\epsilon^2 \, , \\
&\bar F^{\rm \ssc EMI}|^{\req{femi}}_{f_{5}(\theta)} \simeq 1 + 12.7 \epsilon^2 +\mathcal{O}(\epsilon^4) \, , & & \bar F^{\rm \ssc EMI}|^{\rm \ssc Mezei}_{f_{5}(\theta)}=  1 + \frac{40}{\pi} \epsilon^2 \simeq 1 + 12.7\epsilon^2 \, , \\
&\bar F^{\rm \ssc EMI}|^{\req{femi}}_{f_{6}(\theta)} \simeq 1 + 22.2 \epsilon^2 +\mathcal{O}(\epsilon^4) \, , & & \bar  F^{\rm \ssc EMI}|^{\rm \ssc Mezei}_{f_{6}(\theta)}=  1 + \frac{70}{\pi} \epsilon^2 \simeq 1 + 22.3\epsilon^2 \, , 
 \end{align}
 %where  we omitted a normalization factor $2\pi^2 \kappa_{(3)}$ in all expressions in order to avoid the clutter. 
As we can see, the numerics do an excellent job in reproducing the exact coefficients in all cases.  A similar analysis in the case of the other family of basis functions, $\sin (\ell \theta)/\sqrt{\pi}$, displays an analogous match with the exact formula for the quadratic piece. We are therefore confident that we can trust our numerical approach and proceed to study regions with non-perturbative shapes.

%\comment{For instance, for the family,
%\begin{equation}
%f(\theta_i)=R[1+\frac{ \epsilon}{\sqrt{\pi}} \cos (\ell \theta_i)]\, , \quad  \text{one gets} \quad
%F^{\rm \ssc EMI}/\kappa_{(3)}=2\pi^2+  \, \frac{2 \pi \epsilon^2}{3}  \ell (\ell^2-1)+\mathcal{O}(\epsilon^3)\, .
% \end{equation}}

\subsection{More general shapes}

Let us move on and consider now more general shapes, which do not correspond to small perturbations of a disk. As anticipated earlier, once we specify the entangling surface via $f(\theta)$, we perform numerical integrations of \req{femi} and \req{femi0} and extract $F^{\rm \ssc EMI}$ from the $\delta \rightarrow 0$ limit. 
It is not obvious at first sight which regions will have a greater $F$. In particular, as mentioned earlier, shapes related by conformal transformations share the same $F$. We do know, nevertheless, that shapes including thin portions will tend to increase $F$ ---see  \req{strip}--- and that shapes which are similar enough to disks will have small $F$'s. Note though that disks deformed by relatively small bumps can modify $F$ considerably, as shown in Fig.\,\ref{refiss376}.

%\begin{figure}[t!] \centering
%	\includegraphics[scale=0.8]{secondfam.pdf}\vspace{-0.7cm}
%	\includegraphics[scale=0.75]{femi2.pdf}
	%\includegraphics[scale=0.63]{elipfamR.pdf}
	%\includegraphics[scale=0.607]{PAab.pdf}
	%\includegraphics[scale=0.45]{ellipses.pdf}\hspace{0.5cm}
	%\includegraphics[scale=0.51]{b15shape.pdf}\hspace{0.5cm}\\
	%\includegraphics[scale=0.51]{b2shape.pdf}\hspace{0.5cm}
	%\includegraphics[scale=0.51]{b25shape.pdf}\hspace{0.5cm}
	%\includegraphics[scale=0.51]{b5shape.pdf}\hspace{0.5cm}
	%\includegraphics[scale=0.5]{circultripa.pdf}
	%\includegraphics[scale=0.71]{stausubal2.pdf}
%	\caption{(Top) We plot various  entangling regions corresponding to $f(\theta)=1/\sqrt{1-e^2 \cos^2 (b \theta})$ for $b=1/2,1/2,3,4,5$ (left to right plots). In each plot, the different figures correspond to different values of $ e \in (0,1)$. (Bottom) We plot the entanglement entropy universal coefficient $F^{\rm \ssc EMI}$, normalized by the disk value, as a function of $e$ for various values of $b$.} %ellipses parametrized by the function $f(\theta)=1/\sqrt{1-e^2 \cos^2 \theta}$ for different values of the eccentricity, $e=0,0.4,0.6,0.8,0.9,0.95,0.99$. (Bottom)}  %(Lower row) We plot elliptic entangling regions characterized by $f(\theta)=a$, $g(\theta)=b$ for $b=1$ and $a=1,3/2,2,5/2,3,7/2$.   }
%	\label{refiss2}
%\end{figure}

\begin{figure}[t] \centering
	\includegraphics[scale=1.23]{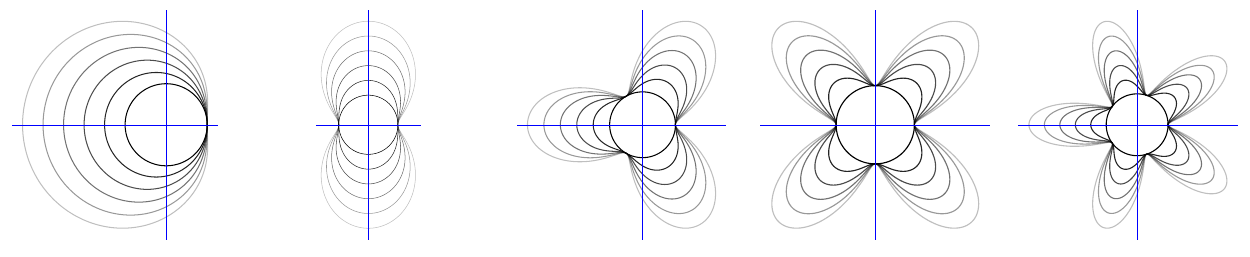}\\ %\vspace{-0.7cm}	
	\includegraphics[scale=0.81]{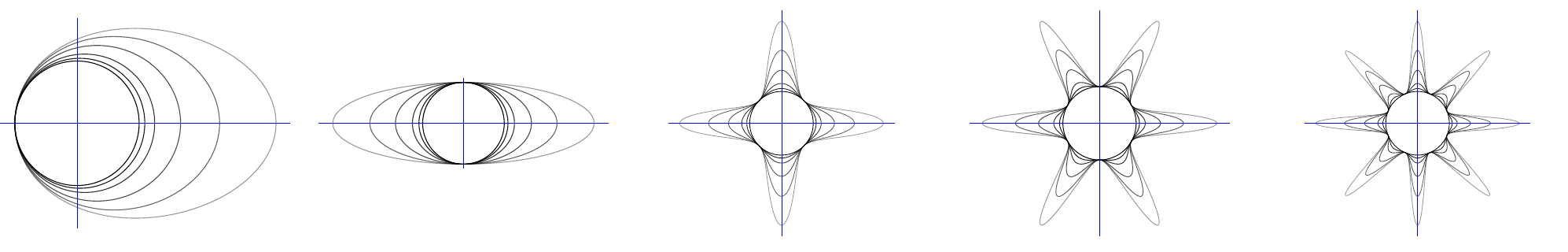}
	%\vspace{-0.7cm}
	\includegraphics[scale=0.673]{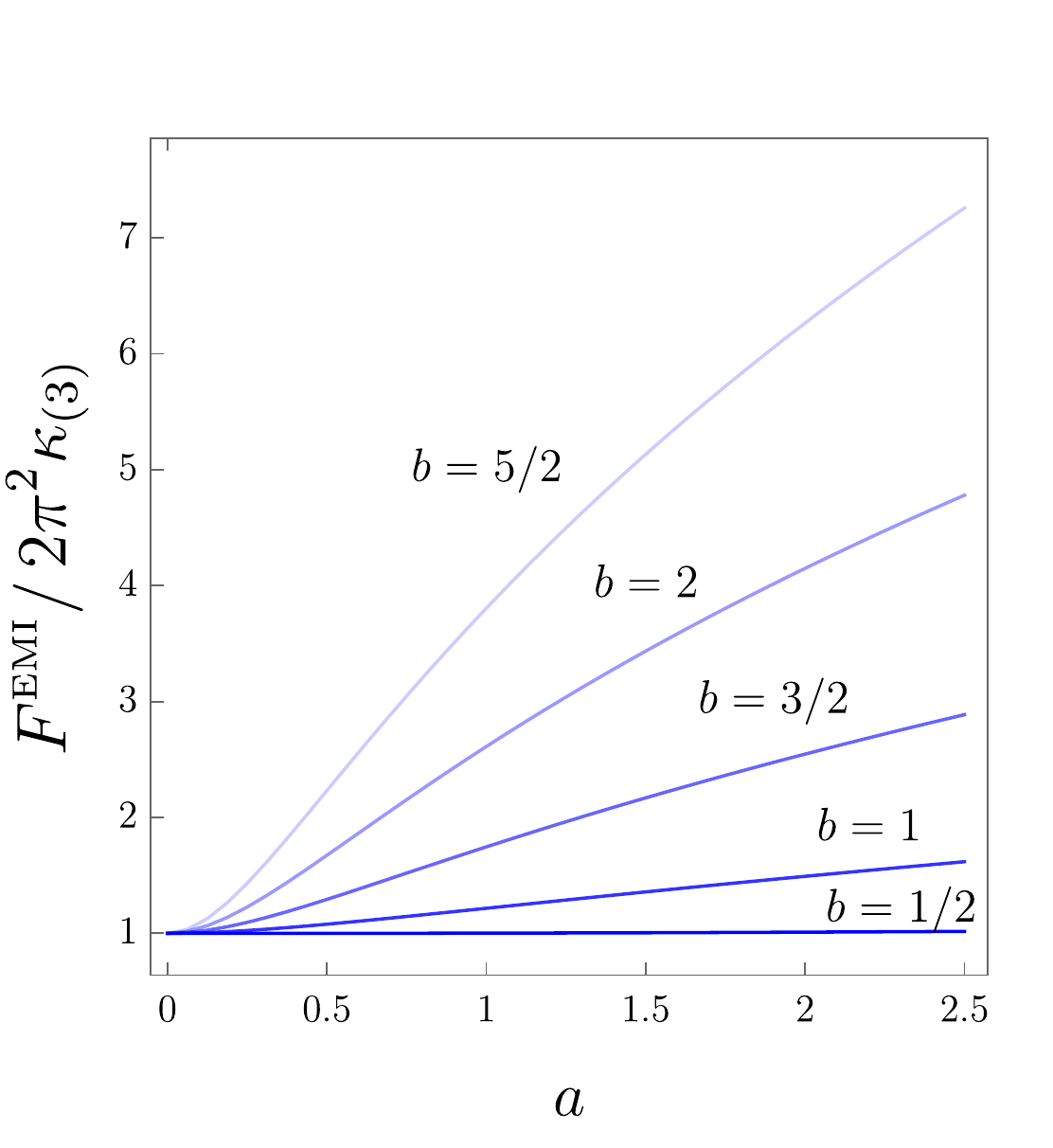}\hspace{-0.2cm}
	\includegraphics[scale=0.67]{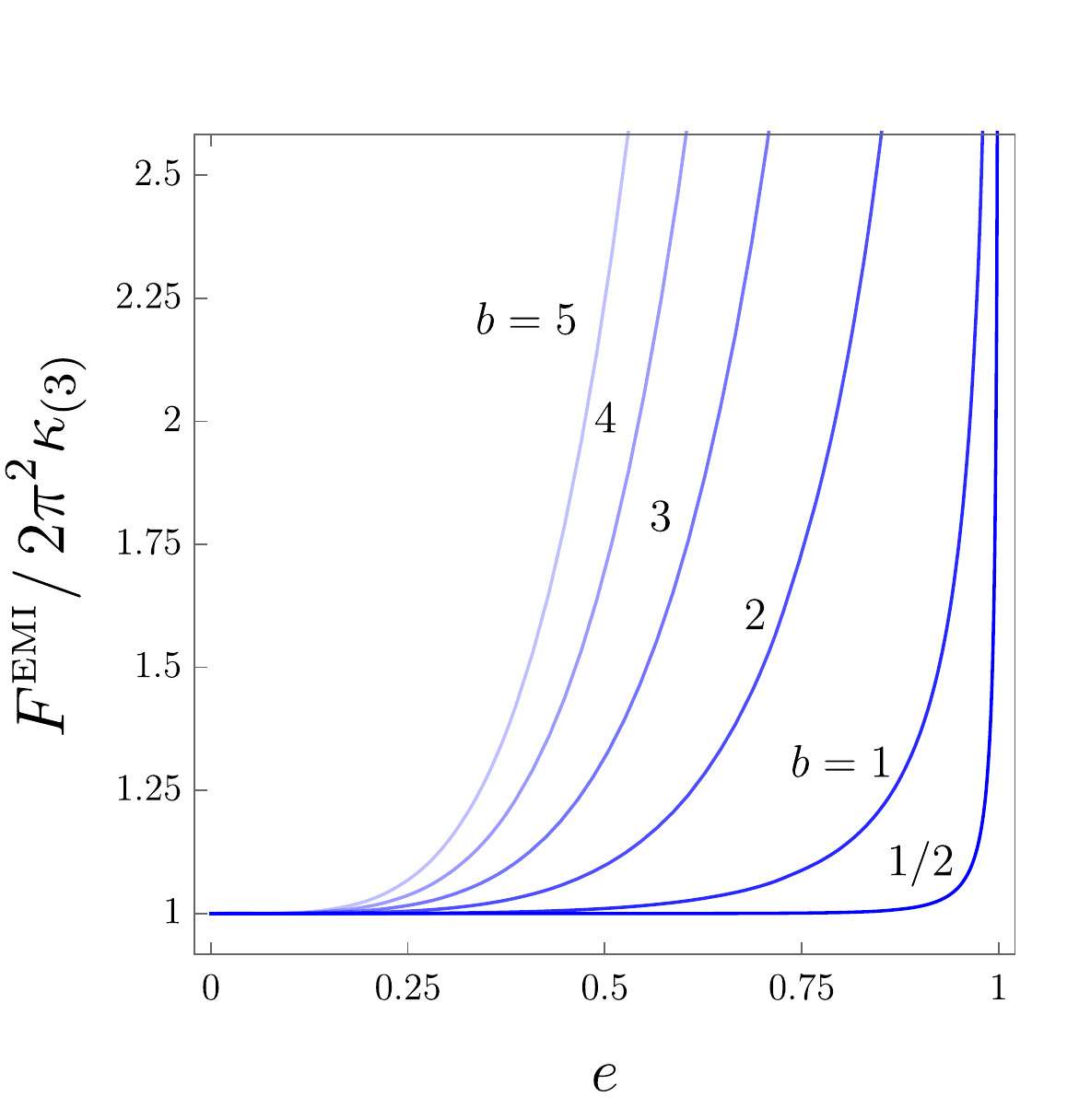}
	\caption{ \textsf{(First row) We plot entangling regions parametrized by the function $f(\theta)=1 + a \sin^2 (b \theta)$ for different values of $a,b$. From left to right we plot five figures corresponding to regions with $b=1/2,1,3/2,2,5/2$ respectively. In each figure, the value $a=0$ corresponds to the disk, and moving outwards we have regions corresponding to $a=1/2,1,3/2,2,5/2$. (Second row) We plot entangling regions corresponding to $f(\theta)=1/\sqrt{1-e^2 \cos^2 (b \theta})$ for $b=1/2,1/2,3,4,5$ (left to right plots). In each plot, the different figures correspond to different values of $ e \in (0,1)$.  (Third row) We plot the entanglement entropy universal coefficient $F^{\rm \ssc EMI}$, normalized by the disk value, as a function of $a$ and $e$, respectively, for various values of $b$ for each family.}} %(Bottom right) We plot the entanglement entropy universal coefficient $F^{\rm \ssc EMI}$, normalized by the disk value, as a function of $e$ for various values of $b$. } %In the right, we plot the analogous curves corresponding to the ratio $\text{perim}^2/(4\pi \, \text{area})$.}  %(Lower row) We plot elliptic entangling regions characterized by $f(\theta)=a$, $g(\theta)=b$ for $b=1$ and $a=1,3/2,2,5/2,3,7/2$.   }
	\label{refiss}
\end{figure}

As our first family, we consider regions defined by the equation
\begin{equation}\label{fami1}
f(\theta)=1+a \sin^2(b \theta)\, ,
\end{equation}
for various values of $a,b$. We plot the ones corresponding to $a=1/2,1,\dots,5/2$, $b=1/2,1\dots,5/2$ in the upper row of Fig. \ref{refiss}. In the same figure, we show the results obtained for $F^{\ssc \rm EMI}$ for half-integer values of $b$ as a function of $a$. We observe that in all cases, the curves lie above the disk result. The values tend to increase as the figures include more geometric features ---\eg as the number of ``petals'' increases. %We also see that the behavior of $F^{\ssc \rm EMI}$ as a  function of $a$ for the different $b$'s is remarkably similar to the one of $\mathcal{R}$ (even though they are clearly different functions).

The second family we consider corresponds to functions
\begin{equation}\label{fami2}
f(\theta)=\frac{1}{\sqrt{1-e^2 \cos^2 (b \theta})}\, ,
\end{equation}
which includes ellipses of eccentricity $e$ as particular cases for $b=1$. We plot the results for $F^{\rm \ssc EMI}$  for various values of $b$ as a function of $e$ in the right plot of the lower row of Fig.\,\ref{refiss}. Once again, we find that all shapes produce results which lie above the disk result and which tend to grow monotonically as the parameters make them become increasingly different from it.  %Similarly to the previous family, we also see that the $F^{\rm \ssc EMI}$ functional displays a behavior analogous to $\mathcal{R}$ as a function of the parameters $e$, $b$. 

Using the ellipse results and the formulas obtained in Section \ref{elipsq}, we can perform another check of the validity of the numerics beyond the perturbative level. In particular, we know that for sufficiently squashed ellipses, the result should approach \req{elie1} for general CFTs. In the case of the EMI model, the strip coefficient $k^{(3)}$ can be computed analytically, and the result reads 
\begin{equation}\label{kemi}
k^{(3)}_{\rm \ssc EMI}= 2\pi \kappa_{(3)}\, .
\end{equation}
For $e=0.99$ one finds $F^{\rm \ssc EMI}|^{ \req{elie1}}_{e=0.99}\simeq 71.9 \kappa_{(3)} $ whereas our numerics give $F^{\rm \ssc EMI}|_{e=0.99}^{\req{femi}} = 71.3 \kappa_{(3)} $, which is already very close ($\sim 0.8\%$ off). The match improves as $e$ grows. For instance, we have $F^{\rm \ssc EMI}|_{e=0.999}^{\req{elie1}} \simeq 221.6 \kappa_{(3)} $ and $F^{\rm \ssc EMI}|_{e=0.999}^{\req{femi}} = 221.2 \kappa_{(3)} $; $F^{\rm \ssc EMI}|^{\req{elie1}}_{e=0.9999} \simeq 698.2 \kappa_{(3)}$ and  $F^{\rm \ssc EMI}|_{e=0.9999}^{\req{femi}} = 697.9$, which differ by  $\sim 0.2\%$ and  $\sim 0.05\%$ respectively. %This level of agreement is in line with the one found between these numerical results and the trial function proposed in \req{trialf}, as shown in the left plot of Fig.\,\ref{refiss254} above.

\begin{figure}[t!] \centering
	\includegraphics[scale=0.6]{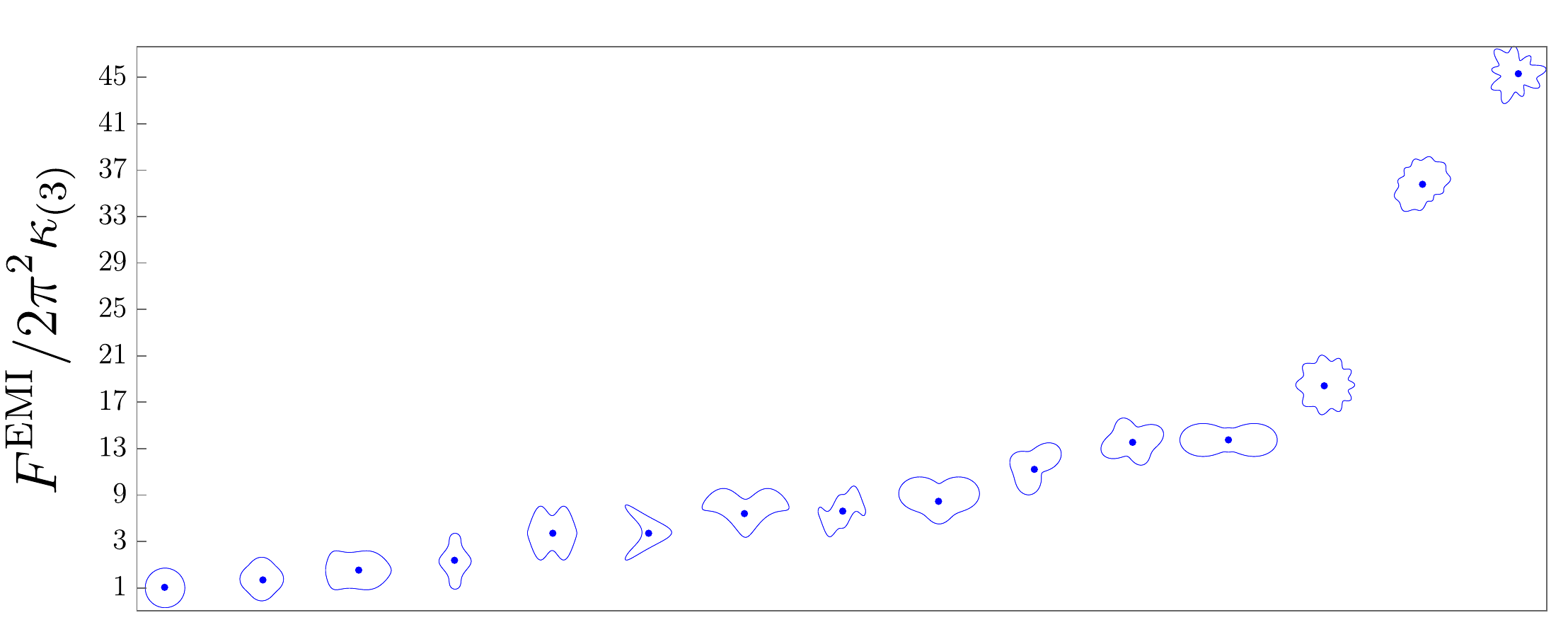}
	\caption{ \textsf{We plot the entanglement entropy universal term normalized by the disk result for the EMI model for various entangling regions. Each data point corresponds to the entangling region displayed on top of it. The shapes are arranged in a way such that $F^{\rm \ssc EMI}$ grows as we move to the right. } }
	\label{refissss2s}
\end{figure}

Naturally, there is a priori no limit in the complexity of the shapes we can probe. We have considered a variety of less symmetric regions and gathered some of the results for $F^{\rm \ssc EMI}$ in Fig.\,\ref{refissss2s}.
 As we can see, the values can vary considerably from shape to shape in a way which is far from obvious by looking at the geometry of the figures. 
 
 As an approximate guiding rule, it turns out that we can establish an analogy with the quantity 
\begin{equation}
\mathcal{R}(A) \equiv  \frac{ \text{perimeter}(\partial A)^2}{4\pi \cdot \text{area}(A)}\, .\label{isopk}
\end{equation}
Similarly to the property we wish to test for $F$, this geometric quantity satisfies  the isoperimetric inequality,
\begin{equation}\label{isopp}
\mathcal{R} \geq 1\, , \quad  \text{with} \quad \mathcal{R}=1 \Leftrightarrow  A = \text{disk}\, .
\end{equation}
Besides, for small deformations of a disk and for very thin strips it behaves, respectively, as
\begin{equation}\label{Rpert}
\mathcal{R}=1+\frac{\epsilon^2}{2\pi}\sum_{\ell\geq 2} \left(\ell^2-1\right)\left[ (a^{(c)}_{\ell})^2+(a^{(s)}_{\ell})^2 \right]+\mathcal{O}(\epsilon^4)\, , \quad \mathcal{R}\simeq  \frac{L}{4\pi r}+ \dots ,
\end{equation}
which are rather similar to the general expressions for $F$ in those regimes. Hence, it is natural to wonder about possible relations between the two quantities. Obviously, while $\mathcal{R}$ is a fixed number for a given $A$, $F$ depends on the theory under consideration, so one can only expect the analogy to be approximate at best. Besides, as opposed to $F$, $\mathcal{R}$ is not conformally invariant. In the case of the EMI model, we find that in most cases we have tested, given two entangling regions $A$, $B$ such that $\mathcal{R}_A > \mathcal{R}_B$, it happens that $F^{\rm \ssc EMI}_A> F^{\rm \ssc EMI}_B$ and viceversa. However, this is not a general rule, and we have found counterexamples. For instance, if $A$ is an ellipse of eccentricity $e=0.97$ and $B$ is a region defined by $f(\theta)=1+15/26\cdot \sin(2\theta)^2$, we have $\mathcal{R}_A/\mathcal{R}_B\simeq 1.12514$ whereas $F^{\rm \ssc EMI}_A/ F^{\rm \ssc EMI}_B \simeq 0.871263$.

In sum, we observe that the explicit evaluation of $F$ in a concrete model produces results which are always larger than the disk one and which can vary very considerably as we change the entangling region, even without the need of considering shapes with very thin sectors.

\section{General proof using strong subadditivity}\label{proof}
In this section we provide a general proof for \req{Fiso}. Namely, we establish that $F$ is globally minimized by disk regions for general theories. Before getting there we require some preliminary results regarding general inequalities satisfied by $F$ which follow from the  strong subadditivity of EE and the behavior of $F$ under different kinds of geometric deformations for a given entangling region. These are included in the first two subsections. The proof is presented in subsection \ref{proooof}.  %In the first subsection we comment on a couple of general identities satisfied by $F$ which follow from the strong subadditivity of EE and which are used in the proof. Then, we analyze the type of deformations which allow for perturbative expansions of 
%The entropy of the vacuum for a region $A$ with smooth boundary in a $d=3$ QFT has the form
%\be
%\see(A)=c\,\frac{P(A)}{\delta}-F(A)\,.\label{entro}
%\ee

\subsection{Strong superadditivity of $F$}
 
 The entanglement entropy of two entangling regions  $\gamma_A$ and $\gamma_B$  satisfies the strong subadditivity property\footnote{In this section, we will indistinctly denote entangling regions and their boundaries by $\gamma$. This will avoid some unnecessary clutter in the expressions.   }
\be\label{ssaEE}
\see (\gamma_A)+\see (\gamma_B)\ge \see (\gamma_{A}\cap \gamma_{B})+\see (\gamma_{B}\cup \gamma_{B})\, .
\ee
In the case of three-dimensional QFTs, this implies for the universal term
%This implies 
\be
F(\gamma_A)+F(\gamma_B)\le F(\gamma_{A}\cap \gamma_{ B})+F(\gamma_{A}\cup \gamma_{B})\,.\label{ine}
\ee
This follows from \req{ssaEE} because the perimeters cancel in the combination. We call this property of $F$ ``strong supperadditivity'' (SSA), having the opposite sign to strong subadditivity because of the conventional sign of $F$ in (\ref{entro}). The inequality (\ref{ine}) holds even if the intersection and union do not have smooth boundaries. If these boundaries have discontinuous first derivatives, the difference between the two sides of the inequality is in fact infinite. As mentioned earlier, here we will need only be concerned with smooth intersections and unions.    

For pure states, the entropy for a region $\gamma$ is equal to the one of its complement $\bar \gamma$, and this implies 
\be
F(\gamma)=F(\bar \gamma)\,. \label{bar}
\ee
Eqs. (\ref{ine}) and (\ref{bar}) together give rise to a new inequality which reads
\be
 F(\gamma_A)+F(\gamma_B)\le F(\gamma_{A} -\gamma_B)+F(\gamma_{B}- \gamma_A)\,,\label{ine2}
\ee
where the difference $\gamma_A-\gamma_B$ stands for the relative complement $\gamma_A\cap \overline{\gamma_B}$. 

We will consider regions lying in the plane $t=0$. In this case the inequality  (\ref{ine}) cannot be nontrivially saturated in QFT. This is a consequence of the Reeh-Schlieder property.  The saturation of (\ref{ine}) is called the Markov property and only occurs in special situations such as regions with boundary in the null plane, or in CFTs for regions with boundaries in the null cone \cite{Casini:2017roe,Casini:2017vbe}.

 \subsection{Geometric perturbations of $F$ and $4$-expandable functionals}

%\section{F is minimal for circles}
In order to show that the property of SSA implies that $F$ is globally minimized for circles in a CFT,  a preliminary step concerns the behavior of the functional $F$ under small perturbations.
%From now on, we will also denote the boundary of a given region by $\gamma$.
%To simplify the notation we will call the boundary of a region $\gamma$ with the same name.
 We consider only smooth boundaries for entangling regions and call ${\cal S}$ the space of curves in the plane which are boundaries of compact regions with smooth and finite boundary. Let $s$ be a length parameter along $\gamma$, and call $\eta(s)$ to the outward pointing normal unit vector.  We call a perturbation of $\gamma \in {\cal S}$ to a set of curves $\gamma_\epsilon \in {\cal S}$, $\epsilon\in[0,\epsilon_0]$, given by 
\be
\gamma_\epsilon(s)=\gamma(s)+\delta_\epsilon(s)\,\eta(s)\,, 
\ee
where the smooth function $\delta_\epsilon(s)$ satisfies $\lVert \delta_\epsilon \rVert\le \epsilon$, and where we write the uniform norm $\lVert h \rVert =\max_{s} |h(s)|$.  If the perturbation is given by 
\be
\delta_\epsilon(s)=\epsilon\, h(s)\,,
\ee
for a fixed function $h(s)$, such that the function with all its derivatives go to zero with $\epsilon$ with the same velocity, we should have a power expansion for $F$ of the form
\be
F(\gamma_\epsilon)=  F(\gamma)+ \int \diff s\, A_1^\gamma(s)\,\delta_\epsilon(s)+\int \diff s_1\, \diff s_2\, A_2^\gamma(s_1,s_2) \,\delta_\epsilon(s_1)\,\delta_\epsilon(s_2)+{\cal O}(\epsilon^3)\,.   \label{expa0}  
\ee
The question we want to address is what happens for more general perturbations where the derivatives of $\delta_\epsilon(s)$ do not go to zero with the same velocity, or even  do not go to zero at all.   

 Physically, only ultraviolet entanglement can be sensitive to the derivatives of $\delta_\epsilon$ as the perturbation size goes to zero. Then, it is clear that the non-local part of the functional $F$, which relates the perturbation at a point with the shape of the curve or the perturbation far away, must still satisfy (\ref{expa0}) with the first two terms going to zero as $\epsilon$ and $\epsilon^2$ respectively. To understand local contributions due to short length entanglement we can forget about the global shape of $\gamma$. These local terms are included for example in the distributional nature of the kernel $A_2^\gamma(s_1,s_2)$ in the vicinity of coincidence points. Locally, scale invariance implies (for dimensional reasons) the form
\be
\int \diff s_1\, \diff s_2\, \frac{\delta_\epsilon(s_1)\,\delta_\epsilon(s_2)}{|s_1-s_2|^4} \sim \int \diff s_1\, \diff s_2\, \log(|s_1-s_2|)\,\delta_\epsilon''(s_1)\,\delta_\epsilon''(s_2)\, ,
\ee  
for the most singular possible term for the contribution quadratic in the perturbation. 
%This is the most singular possible term for the contribution quadratic in the perturbation.
 Powers of the local curvature have positive dimension and can only soften the contribution.    The definition of the regularization of the distribution on the left hand side of the equation is given by the right hand side.

In momentum space the above term writes for large momentum   
\be
\int \diff p \, |\tilde{\delta}_\epsilon(p)|^2\, p^3\,,
\ee 
which is precisely the form of the leading angular momentum term for deformations of disk regions   ---see eq. \req{fmeze0}. The coefficient of this singular term must be independent of the precise form of $\gamma$ and is proportional to $C_{\ssc T}$ \cite{Mezei:2014zla}.  For comparison, the perimeter functional, which is not scale invariant, has the milder leading local term
\be
\int \diff p \,  |\tilde{\delta}_\epsilon(p)|^2\, p^2\,.
\ee
This has dimensions of length since $\tilde{\delta}_\epsilon$ has length-square dimensions.

Analogously, we can analyze the behavior of possible higher non-linear terms in the perturbation. In the scale-invariant case, in momentum space we have a leading behavior for large momentum given by  %\comment{this is the leading behavior for large momentum of what?}
\be
\int  \diff p\, |\tilde{\delta}_\epsilon(p)|^{n} p^{2 n-1}\,.\label{lab}
\ee   

For the sake of the proof below, we are interested in understanding perturbations of the form $\delta_\epsilon(s)=\lambda^{a}\, h(s/\lambda)$, where we have written $\epsilon=\lambda^a$, $a>0$. For these, the derivatives go as $\delta_\epsilon^{(k)}\sim \lambda^{a-k}$. In momentum space, $\tilde{\delta}_\epsilon(p)\sim \lambda^{a+1}$, $p\sim  \lambda^{-1}$. The leading local term of $n^{\rm th}$ order (\ref{lab}) goes as $\lambda^{n (a-1)}$. On the other hand, the non-local pieces scale as $\lambda^{(a+1) n}$.     
For these perturbations, in order to separate the first term ($n=1$) of (\ref{expa0}), from the second ($n=2$), we need $a>3$.\footnote{\rd{For the $n=1$ and $n=2$ terms we have for the non-local and leading local pieces, respectively: $\{ \lambda^4,\lambda^2\}$ and $\{ \lambda^8,\lambda^4\}$ for $a=3$;  $\{ \lambda^5,\lambda^4\}$ and $\{ \lambda^{10},\lambda^6\}$ for $a=4$. Hence, using $a=3$ would mix the non-local $n=1$ contribution with the local $n=2$ one. }}  We will use $a=4$ in the proof. For $a=4$, the first term in (\ref{expa0}) is order $\lambda^5$, while the second (the leading local piece) is order $\lambda^6$. The rest of the contributions start at order $\lambda^7$.  This motivates the following definition. 

\bigskip

\noindent {\bf Definition:} { \sl We call a functional $f:{\cal S}\rightarrow \mathbb{R}$ ``a-expandable'' if for any perturbation $\delta_\lambda(s)$ of any $\gamma\in {\cal S}$, such that $\lVert \delta_\lambda^{(k)}\rVert \sim \lambda^{a-k}$ as $\lambda \rightarrow 0$, the following expansion is valid
\be
f(\gamma_\lambda)=  f(\gamma)+ \int \diff s\, A_1^\gamma(s)\,\delta_\lambda(s)+\int \diff s_1\, \diff s_2\, A_2^\gamma(s_1,s_2) \,\delta_\lambda(s_1)\,\delta_\lambda(s_2)+\dots \, ,    \label{expa01}  
\ee
where the second-order term is higher order in $\lambda$ than the linear term, and the rest is higher order than the second term.  } \bigskip

Naturally, $F$ is an example of a $4$-expandable functional.

 \subsection{$F$ is globally minimized by disk regions} \label{proooof}

With this understanding we are in position to prove the minimality of $F$ for disk regions. In order to highlight the geometric nature of the proof we state it for general functionals. 

\bigskip

\noindent {\bf Theorem:} {\sl If an Euclidean invariant\footnote{Namely, invariant under rotations and translations.} 4-expandable functional $F$ on ${\cal S}$ is  strong superadditive and constant for disks, then $F$ is globally minimized for disks.}

\bigskip 

{\sl Proof}:  First we deal with regions with non-trivial topology. Suppose $\gamma$ has more than one connected components $\gamma=\gamma_1\cup \cdots \cup\gamma_n$. From SSA 
\be
F(\gamma)\ge \sum_{i=1}^n  F(\gamma_i)\,.  \label{sapa}
\ee
This shows that multicomponent regions have larger $F$ than certain single-component ones. Then suppose $\gamma$ is a single-component region with holes, $\gamma=\gamma_0-(\gamma_1\cup \dots\cup \gamma_n)$, where $\gamma_0,\gamma_1\,\cdots,\gamma_n$ are single-component simply connected regions. From (\ref{ine2}) it follows
\be \label{514}
F(\gamma) \ge  F(\gamma_0) +\sum_{i=1}^n  F(\gamma_i)\,,
\ee
which shows that $F(\gamma)$ is greater or equal than some single-component region without holes. Therefore, in order to prove the theorem, we can from now on restrict ourselves to single-component simply connected regions.

 Consider then a region $\gamma$ and another region $\tilde{\gamma}$ included in it, $\tilde{\gamma}\subseteq \gamma$. Assume $\tilde{\gamma}$ is osculating to $\gamma$, that is, both curves are tangent to each other and at the point of contact ---which we can set to the length parameter $s=0$ for both curves--- their curvatures are equal, $\tilde{\gamma}''(0)=\gamma''(0)$. Since one curve is included inside the other, the third derivatives must also agree $\tilde{\gamma}'''(0)=\gamma'''(0)$, and the deviation between them will be fourth order $|\tilde{\gamma}(s)-\gamma(s)|\sim |s|^4$ near $|s|=0$. 
 
 \begin{figure}[t!] \hspace{-0.35cm}
	\includegraphics[scale=0.45]{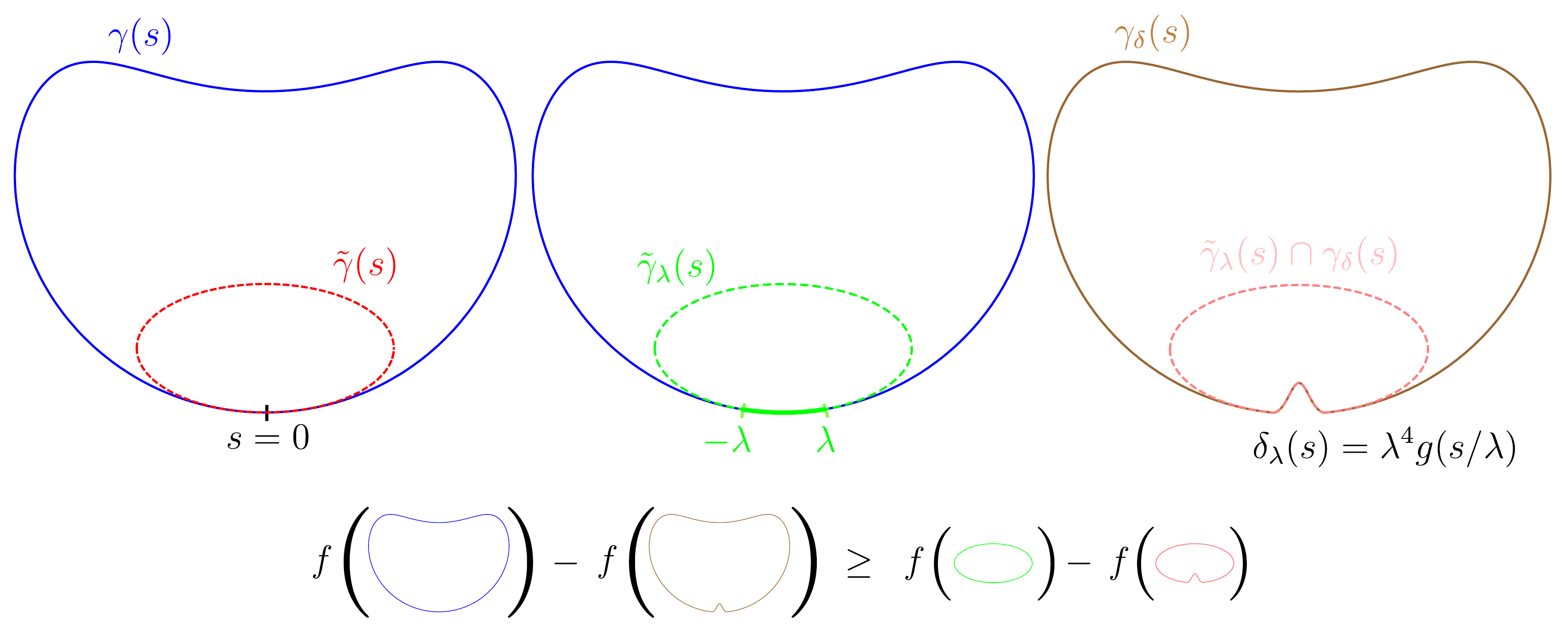}
	\caption{ \textsf{We  show an schematic representation of the geometric setup leading to the inequality \req{dsf}. In the left plot we show two entangling regions, $\tilde \gamma(s) \subseteq \gamma(s)$ such that $\tilde \gamma(s)$ is osculating to $\gamma(s)$ at $s=0$. The second plot includes $\gamma(s)$ and a deformed version of the inner region, $\tilde \gamma_{\lambda}(s)$, which exactly coincides with $\gamma(s)$ for  $s\in (-\lambda,\lambda)$. The third plot shows deformed versions of $\gamma(s)$ and $\tilde \gamma_{\lambda}(s)$ which include a small inwards bump supported within the interval for which both regions coincide. The equation shown below is just \req{fgg} in the particular case considered above.  }}
	\label{refides}
\end{figure}
 
 Around the point $s=0$ we consider a family of deformations $\tilde{\gamma}_\lambda(s)$ of $\tilde{\gamma}$, with deformation function $\tilde{\delta}_\lambda(s)$, such that the deformation has support in $s\in(-2\lambda,2 \lambda)$, and the curve $\tilde{\gamma}_\lambda(s)$ coincides with $\gamma(s)$ for $s\in (-\lambda,\lambda)$. Then, the deformation can be chosen such that $\lVert\tilde{\delta}_\lambda^{(k)}\rVert\sim  \lambda^{4-k}$. With support inside the interval $s\in (-\lambda,\lambda)$ where $\tilde{\gamma}_\lambda$ and $\gamma$ coincide, we can place a deformation of $\gamma$ given by $
\delta_\lambda(s)=\lambda^4\, g(s / \lambda)$, 
where $g$ is a smooth function. We also impose $g< 0$ such the deformation is inwards ---see Fig.\,\ref{refides} for a representation of the geometric setup just described. We also have $\lVert\delta_\lambda^{(k)}\rVert \sim \lambda^{4-k}$. 
 From SSA we find then\footnote{\rd{In particular, \req{fgg} follows from choosing $\gamma_A\equiv \gamma_{\delta}$ and $\gamma_B=\tilde{\gamma}_{\lambda}$ in  \req{ine}.}}
\be \label{fgg}
F(\gamma)-F(\gamma_\delta)\ge F(\tilde{\gamma}_\lambda)-F(\tilde{\gamma}_\lambda\cap \gamma_\delta)\,.
\ee 
Expanding for small $\lambda$, we have for the first-order deviation
\be
A_1^{\gamma}(0)-A_1^{\tilde{\gamma}}(0)\ge 0\,. \label{dsf}
\ee
That is, to first order, $f$ for the larger region increases more than for the smaller one going outwards at the point of osculation.
 
  For a disk region and an Euclidean invariant $F$, the coefficient $A_1^{\rm disk}(s)$ has to be independent of $s$. Then, taking a first order variation from a disk to a scaled disk, and considering that $F$ is constant on disks, it follows that 
  \be
  A_1^{\rm disk}(s)=0\,.\label{circlea}
\ee  
   This will be useful when combined with \req{dsf}. In order to use this result, we have first to discuss some geometric preliminaries. We will show that for an arbitrary curve $\gamma$ we can always place an osculating circle inside (or outside) it.    
 
\begin{figure}[t!] \centering
	\includegraphics[scale=0.5]{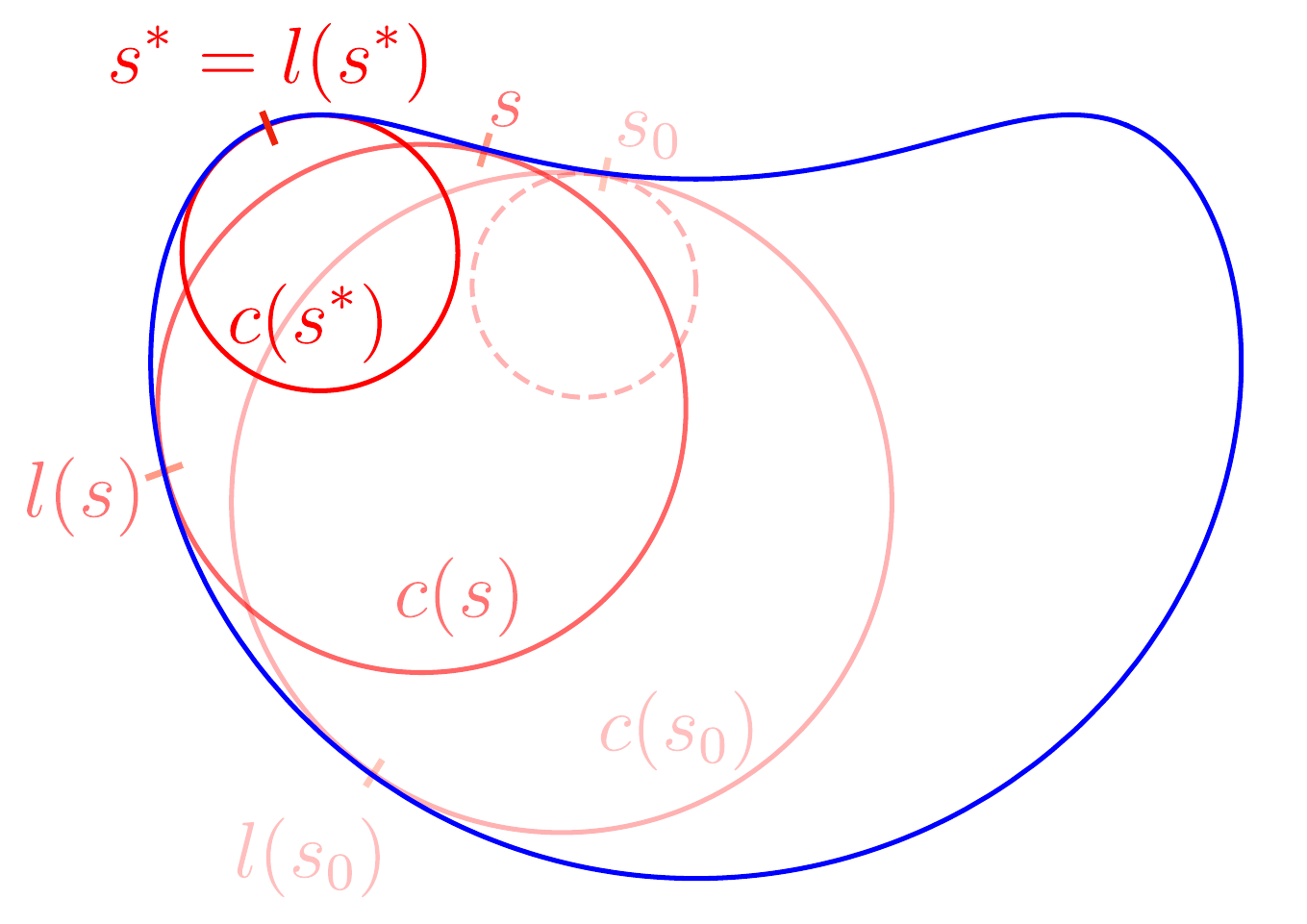}
	\caption{ \textsf{We show the procedure outlined in the text for finding an osculating circle $c(s^*)$ inside a given shape. }}
	\label{refidesss8}
\end{figure}

 To show this, call the length parameter $s\in [0,L]$. Take a point $s_0$ and a small circle internal to $\gamma$ and tangent to $\gamma$ at $s_0$ ---see Fig.\,\ref{refidesss8}. By increasing the radius of this circle and keeping it tangent to $\gamma$ at $s_0$ we arrive to a unique circle $c(s_0)$ which is still included in $\gamma$ and is tangent to $\gamma$ at $s_0$ and at (at least) another point $l(s_0)$. We can choose the length parameter such that $s_0>0$, $l(s_0)<L$, $s_0<l(s_0)$. We define the function $l(s)$ for another $l(s_0)>s>s_0$ in a similar way, by taking the largest circle $c(s)$ tangent at $s$ and internal to $\gamma$, and where $l(s)$ is the smallest length parameter (greater than $s$) among the points at which the circle $c(s)$ is tangent to $\gamma$. It is clear that moving $s$ from $s_0$ to larger values,  $l(s)$ can only decrease because $\gamma([s,l(s)])$ and a segment of the circle $c(s)$ from the point $s$ to $l(s)$ defines a closed curve that divides the plane in two regions. Any $c(s')$ for $s<s'<l(s)$ is included in this region and will have $l(s')< l(s)$. Then, there is a smallest $s^*>s_0$ such that $s^*=l(s^*)$. This indicates that $c(s^*)$ is an internal osculating circle to $\gamma$. At this point the osculating circle and $\gamma$ have a point of contact of degree four, the curvatures of $\gamma$ and $c$ agree, and the derivative of the curvature of $\gamma$ vanishes. In fact $s^*$ is a local maximum of the curvature. If it where a minimum, $\gamma$ would leave part of the circle outside.

{ We have shown that for any $\gamma$ we can place an osculating circle inside it. It is not difficult to show, using the same ideas, that we can also find an external osculating circle to $\gamma$. A shorter proof follows by using conformal transformations. We can make a conformal inversion $I$ centered at a point inside $\gamma$. As a result $\gamma$ is mapped to a closed curve $\tilde{\gamma}$, and the exterior  of $\gamma$ is mapped to the interior of $\tilde{\gamma}$.  Again there will be a  circle $\tilde{c}$ osculating to $\tilde{\gamma}$ internally. Considering that the osculating condition is conformally invariant, and that circumferences are mapped to circumferences, by inverting back to the original plane we obtain that there is a circumference $c=I(\tilde{c})$ osculating to $\gamma$, and this circumference is completely included in the exterior of $\gamma$. The corresponding circle either lies completely outside $\gamma$, in which case the point of contact is a local minimum of curvature for $\gamma$, with negative curvature, or the circle includes $\gamma$ completely, in which case the contact point is a positive curvature local minimum.}\footnote{{The two cases depend on whether the center of inversion lies inside or outside the osculating circle $\tilde{c}$.}}

{ Hence, we can find a point of local curvature maximum and another of local curvature minimum where we can place internal and external osculating circles. By direct application of the above calculation to the curve $\gamma$ and these osculating circles we get the inequality (\ref{dsf}) for the $A_1$ coefficients of both curves. From the equation (\ref{circlea}) for circles,  we find that for any $\gamma$ there is at least one point of local curvature maximum where $A_1\ge 0$ and one point of local curvature minimum where $A_1\le 0$. If $|A_1|>0$, the sign of $A_1$ is kept constant in a finite interval around the point, whereas for $A_1=0$ it can change sign there.  We can use a perturbation of order $\epsilon$, with all derivatives of the same order, to increase the size of the local curvature minimum or decrease the size of the local curvature maximum, by pushing the curve to the outside or the inside respectively.\footnote{{ This can be done, for example, by replacing a small interval of the curve around the point of osculation by a segment of a circle tangent at two points at each side of it, and smearing up the points of contact to make it smooth. The same can be done if the minimum or maximum of curvature correspond to a segment of a circle rather than a point.}}   
If $|A_1|>0$ we can then use the expansion (\ref{expa01}) for this perturbation and check that $F$ always decreases. In the case $A_1=0$ the perturbation can be chosen such that the change in $F$ vanishes to the order of the change in the curvature. 
 For any $\gamma$, we have some curvature maxima $\{r_1^+,\cdots, r_k^+\}$ and minima $\{r_1^-,\cdots,r_k^-\}$, and define $K(\gamma)=\sum_{i=1}^k r_i^+ - \sum_{i=1}^k r_i^-\ge 0$. 
 Then we have shown that for any $\gamma$ there is a $\gamma'$ with $K(\gamma')< K(\gamma)$ and $F(\gamma')\le F(\gamma)$. Hence, the function
 \be
 f(q)=\textrm{min}_{\gamma, K(\gamma)=q}\, F(\gamma) \,,      
 \ee
has to be non decreasing with $q$. In particular, the absolute minimum of $F$ has to be achieved for $q=0$, corresponding to a circle. }  
 $\square$

\bigskip

Therefore, the $F$ term in the entanglement entropy of three-dimensional CFTs is globally minimized for disk regions.
Except for the particular continuity property of the functional that we used, which is natural for conformal invariant functionals, we did not need to invoke conformal invariance but just Euclidean invariance, plus the requirement that the functional is constant for disks. 
The proof should go over to the Lorentzian case as well. In that case, however, all deformations of the circle on its light cone have the same $F$.  

{ Another comment is that having an $a$-expandable functional for $a>4$ and not expandability for $a=4$ (\ie more singular functionals) is not enough for the proof, since we need to match the boundaries of a deformed osculating circle with the curve, which requires perturbations with $a=4$. In principle, lower $a$-expandability (softer functionals) ---such as the isoperimetric ratio ${\cal R}(A)$ defined in eq. (\ref{isopk}), having $3$-expandability--- will be enough. Note however that ${\cal R}(A)$ is not strong superadditive and we cannot use the same proof as for the isoperimetric inequality. If we have $2$-expandability instead, we could modify the proof above taking just tangent circles instead of osculating ones. However, we always have both interior and exterior tangent circles to a given point, and that would imply a vanishing first order coefficient $A_1$ at any point. Then, the only possibility for a SSA functional to be 2-expandable is that it is constant under deformations.}

\section{Final comments}\label{finalc}
%In this paper we have shown that disks globally maximize the EE of three-dimensional CFTs. 
The main results of the paper appear summarized at the end of the introduction. Let us close with some final comments.

%Our general proof makes use of the strong subadditivity of EE, which for a conveniently engineered geometric setup implies a sign for the variation of $F$ as we modify the shape of a given entangling region. We have also obtained new quantitative results for $F$ in the case of generic elliptic regions valid for general CFTs, and cross-checked them against lattice calculations for free scalars and fermions. The geometric EMI model has allowed us to probe more general shapes to get a grasp on how much $F$ may vary as we depart from the disk case. 
As mentioned in the introduction, the analogous four-dimensional problem was studied in \cite{Astaneh:2014uba} \rd{and \cite{Perlmutter:2015vma}}. \rd{The outcome of the study was in that case that the round sphere is the entangling surface which minimizes the logarithmic coefficient of the EE for general theories as long as $g=0,1$ but that, rather surprisingly, for theories with $a>c$ the coefficient could be lowered arbitrarily by considering certain surfaces of increasing genus \cite{Perlmutter:2015vma} ---see the introduction for more comments}. %Since in that case one of the relevant geometric integrals equals the Euler characteristic of the entangling surface, the authors also considered the natural question of which surfaces maximize the EE for a fixed genus . %For instance, for $g=1$, they argued that the extremizing entangling surface corresponds to the so-called  Clifford torus (for which the quotient of radii equals $1/\sqrt{2}$). 
It is then natural to wonder what happens in three-dimensions with the analogous question for entangling regions with a fixed number of holes. In this case, however, the answer turns out to be more trivial. For instance, for annuli regions defined by pairs of radii from the set $\{R_1,R_2,R_3\}$ with $R_1<R_2<R_3$, one has $F(R_3,R_1)< F(R_2,R_1)$ \cite{Nakaguchi:2014pha}. This implies that $F$ will always decrease as we make the inner radius smaller with respect to the outer one, being $F_0$ the limiting result as $R_1/R_3 \rightarrow 0$. In fact, for regions with non-trivial topology, our general bound \req{Fiso} can be improved. Imagine we have a region with $n$ disconnected subregions $\gamma=\cup_j^n \gamma_j $, each one of which has $m_j$ holes, $\gamma_j=\gamma_{j,0}- (\gamma_{j,1} \cup \dots \cup \gamma_{j,m_j})$. Then, using \req{sapa}, \req{514} and \req{Fiso}  we have 
%for a region $\gamma$ with $n_B$ total boundaries ---this may include holes as in \req{514} and/or disconnected subregions as in  \req{sapa}--- % --- $n$ holes of arbitrary shape, $\gamma=\gamma_0-(\gamma_1\cup \dots\cup \gamma_n)$, 
\begin{equation}
F(\gamma) \geq \sum_j^n F(\gamma_j) \geq  \sum_j^n [F(\gamma_{j,0})+\sum_{i=1}^{m_j} F(\gamma_{j,i})] \geq n F_0 + \sum_j^n m_j F_0=n_B F_0\, .
 \end{equation}
%where $\#_{\gamma}$ is the number of boundaries of the entangling regions ---\ie the number of holes plus one. 
This follows from applying $F \geq F_0$ repeatedly to each single-component simply connected piece. In the last equality we rewrote $n+\sum_j^n m_j$ as $n_B$, which is the total number of boundaries of the region.  In words, for a region with  $n_B$ boundaries, $F$ is bounded below not only by the disk result $F_0$, but by $n_B$ times $F_0$. Saturation of the inequality occurs only for purely topological theories, $F^{\rm topo}(\gamma) = n_B F_0^{\rm topo}$.

%\comment{I guess something analogous happens for all possible configurations with non-trivial topology}
%\comment{other topologies, trivial; different from four dimensions. For annuli, $F(R_3,R_1)< F(R_2,R_1)$ for $R_1<R_2<R_3$ \cite{Nakaguchi:2014pha}, disk result in the limit}

It is also natural to wonder about the problem analogous to the one considered here for $d=4+1$ and $d=5+1$ theories. In the latter case, the geometric nature of the universal logarithmic term for smooth entangling regions \cite{Safdi:2012sn,Miao2015a} should make it amenable to an analysis similar to the one  in \cite{Astaneh:2014uba,Perlmutter:2015vma} for the $d=3+1$ case. Note however that in that case the sign of the different geometric contributions weighted by the trace-anomaly coefficients is not obvious ---at least at first sight--- so the situation is trickier.  The $d=4+1$ case is more challenging but methods similar to the ones considered here may be useful. Naturally, in all cases the expectation is that entangling regions bounded by round (hyper)spheres globally maximize the EE. This expectation is supported by partial evidence from holographic theories \cite{Astaneh:2014uba}.
%\comment{some comments about $d=4+1$ and $d=5+1$}

\begin{acknowledgments}   The work of P.B. and H.C. was supported by the Simons foundation through the It From Qubit Simons collaboration. H.C. was also supported by CONICET, CNEA, and Universidad Nacional de Cuyo, Argentina. The work of JM is funded by the Agencia Nacional de Investigaci\'on y Desarrollo (ANID) Scholarship No. 21190234 and by Pontificia Universidad Cat\'olica de Valpara\'iso. JM is also grateful to the QMAP faculty for their hospitality.
\end{acknowledgments}

\appendix
\section{Mutual information regularization of EE in the EMI model}\label{EMIMIregu}
In Section \ref{secemi} we computed $F$ numerically for many kinds of entangling regions in the EMI model. In order to do so, we introduced a small regulator $\delta$ along an auxiliary extra dimension, evaluated $F^{\rm \ssc EMI}(A)$ as a function of $\delta$ and then extracted the $\delta \rightarrow 0$ limit. However, as we explained in Section \ref{elipsq} there is an alternative way of regularizing EE that makes use of the MI of concentric regions and which yields converging results in the lattice. In this appendix we explore this alternative method for the EMI model in the case of elliptic entangling regions for the two geometric setups explained in Section \ref{elipsq}, namely: fixing a constant $\varepsilon$, and forcing the concentric regions to be ellipses. This will allow us to test to what extent we may expect both methods to differ in general when the size of the entangling regions cannot be considered to be extremely larger than the separation between the auxiliary regions ---which is always the case in the lattice. We will also verify the convergence of both methods between themselves and with the $\delta$ method used in the main text.

In the three-dimensional EMI model, the mutual information between two regions $A,B$ in a fixed time slice is given by
\begin{equation}\label{emimi}
I^{\rm \ssc EMI}(A,B)=-2\kappa_{(3)}\int_{\partial A}\diff\mathbf{r}_A\int_{\partial B}\diff\mathbf{r}_B\, \frac{\mathbf{n}_A\cdot\mathbf{n}_B}{\abs{\mathbf{r}_A-\mathbf{r}_B}^{2}}\, ,
\end{equation}
where the normal vectors are defined outwards to each region, respectively. This formula is manifestly non-divergent, as $\mathbf{r}_A$ and $\mathbf{r}_B$ correspond now to different regions. For regions $A,B$ parametrized by $[x_{A,B}(t),y_{A,B}(t)]$ we have
\begin{equation}\label{emimi}
I^{\rm \ssc EMI}(A,B)=-2\kappa_{(3)}\int_{\partial A}\diff t_A\int_{\partial B}\diff t_B\, \frac{ \dot y_A(t_A) \dot y_B(t_B) + \dot x_A(t_A) \dot x_B(t_B)  }{  (x_A(t_A)-x_B(t_B))^2+(y_A(t_A)-y_B(t_B))^2}\, .
\end{equation}

Applied to the setup required for regularizing the EE, as in \req{annulii}, we need to evaluate the mutual information $I^{\rm \ssc EMI}(A^+,A^-)$, which involves integrals over $\partial A^+=\partial \overline{A^+}$ and $\partial A^-$. Using the EE formula for the EMI model, \req{emiee}, it is easy to see that in $I^{\rm \ssc EMI}(A^+,A^-)$ the contributions from $\see^{\rm \ssc EMI}(\overline{A^+})$ and $\see^{\rm \ssc EMI}(A^-)$ cancel with the terms in $\see^{\rm \ssc EMI}(\overline{A^+ \cup A^-})$ which involve performing both integrals over  $\partial A^+$ or both integrals over $\partial A^-$. The only terms which survive are the ones which involve one integral over $\partial A^+$ and the other one over $\partial A^-$, in agreement with \req{emimi}.

 \begin{figure}[t!] \centering
	\includegraphics[scale=0.7]{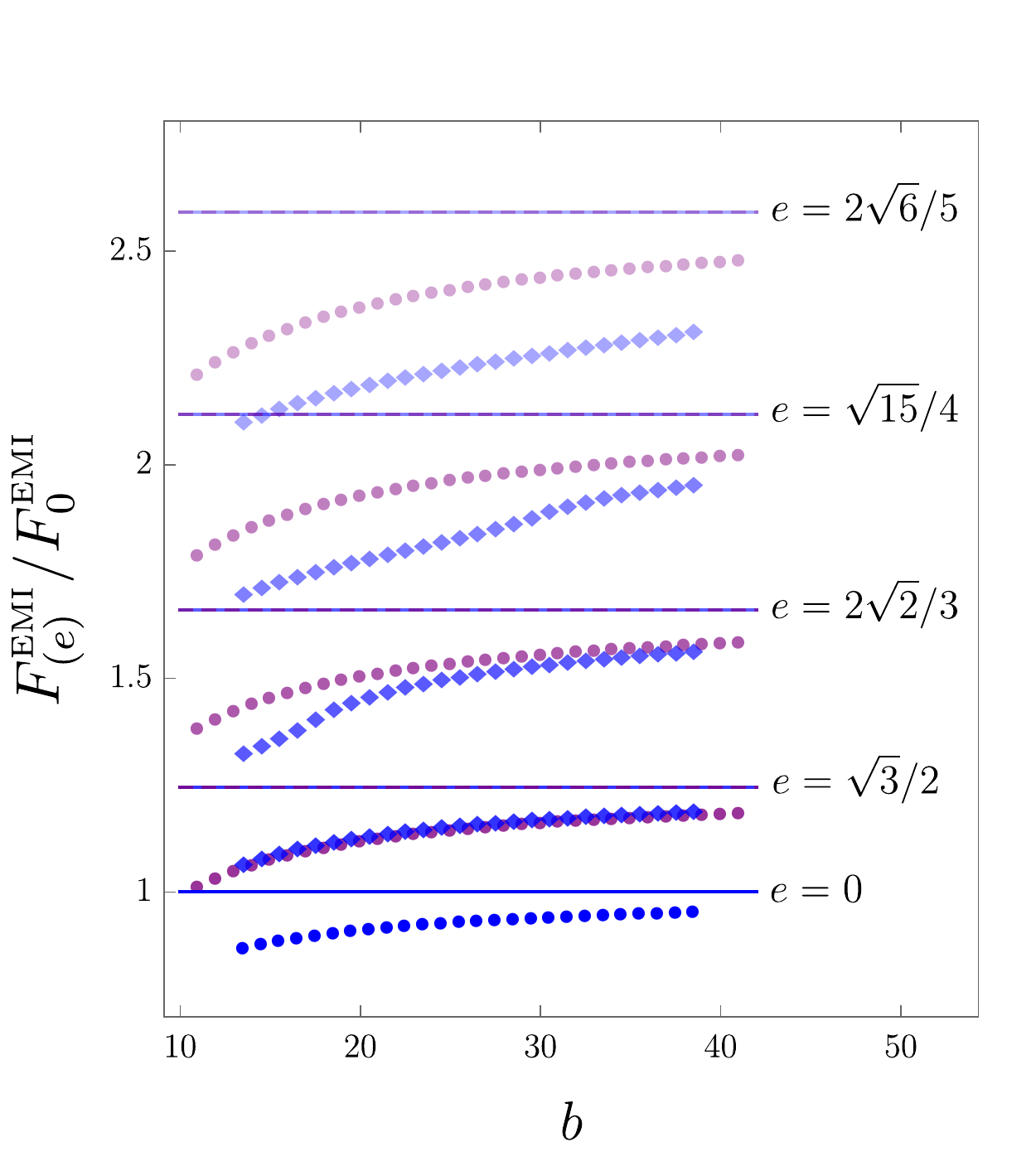}
	\caption{ \textsf{We plot $F_{(e)}^{\rm \ssc EMI}/F_0^{\rm \ssc EMI}$ for different values of the eccentricity $e= 0,\sqrt{3}/2,2\sqrt{2}/3,\sqrt{15}/4,2\sqrt{6}/5$ using the expressions \req{ik130emi} (purple dots) and \req{ik13emi} (blue diamonds). In the case of  \req{ik130emi}, we fix $\varepsilon=5$ and plot the results as a function of the semi-minor axis $b$. In the case of \req{ik13emi} we fix $b_2=b_1+7$ and plot the results as a function of $b=(b_1+b_2)/2$ as well. The straight lines correspond to the exact values obtained using the $\delta$ method, which can be also reproduced using \req{ik130emi} or \req{ik13emi} for sufficiently large values of $b$.}}
	\label{ref44}
\end{figure}

%\subsection{Constant $\varepsilon$}
%The relevant formulas for the two relevant cases, 
In the case in which we keep $\varepsilon$ constant for all points of the ellipse boundary, the relevant formula is
 \begin{equation}\label{ik130emi}
 F^{\rm \ssc EMI}_{(e)}=-\frac{1}{2}\left[  I^{\rm \ssc EMI}({\rm pseudoellipse }_{i},{\rm pseudoellipse}_{o})-k_{{\rm \ssc EMI}}^{(3)} \frac{4 a E[e^2]}{\varepsilon}\right]\, ,
 \end{equation}  
where the parametrization of the pseudoellipses appears in \req{rio} and $k_{{\rm \ssc EMI}}^{(3)}$ is defined in \req{kemi}. On the other hand, if we force the auxiliary regions to be ellipses, the relevant formula reads
 \begin{equation}\label{ik13emi}
 F^{\rm \ssc EMI}_{(e)}=-\frac{1}{2}\left[  I^{{\rm \ssc EMI}}({\rm ellipse }_{2},{\rm ellipse}_{1})-k_{{\rm \ssc EMI}}^{(3)} \int_0^{\frac{\pi}{2}} \diff t \frac{2(a_1+a_2)\sqrt{1-e^2 \cos^2 t} }{\varepsilon(t)}\right]\, ,
 \end{equation} 
where $a=(a_1+a_2)/2$, $b=(b_1+b_2)/2$ and the formula for $\varepsilon(t)$ appears in  \req{epis}.

As a first check, we have verified that both \req{ik130emi} and \req{ik13emi} produce results essentially identical to the ones obtained using the $\delta$ method for sufficiently large values of $b/\varepsilon$ ---\eg $b/\varepsilon \sim 100$ yields excellent results in all cases. On the other hand, practical limitations force us to consider smaller values of such a quotient when performing calculations in the lattice ($b/\varepsilon \sim 10$). In Fig.\,\ref{ref44} we have plotted the results achieved using both methods for values of $b$ similar to the ones considered in our lattice calculations in Section \ref{elipsq}. As we can see, in both cases the results tend to underestimate considerably the actual ones ---this underestimation tends to be greater as the eccentricity grows. Importantly, we observe that the pseudoellipses method makes a better job in approximating the actual results than the variable-$\varepsilon$ one. The difference between both methods becomes rather considerable for greater values of the eccentricity. For each eccentricity and each value of $b$, we can observe what is the factor we need to multiply the corresponding EMI result by in order to obtain the exact answer $F_{(e)}^{\rm \ssc EMI}$. We use such factors in Section \ref{elipsq} to correct the lattice results obtained for the corresponding values of $b/\delta$.

\bibliography{Gravities}
\bibliographystyle{JHEP-2}
\label{biblio}

\end{document}